\documentclass[journal]{IEEEtran}

\usepackage[utf8]{inputenc}

\usepackage{amsmath,amssymb}

\usepackage{xcolor}
\usepackage{graphicx}
\usepackage{tikz}
\usepackage{pgfplots}
\usepackage{float}
\usepackage{balance}

\usepackage{xspace}
\usepackage[normalem]{ulem}
\usepackage[pdftex]{hyperref} 
\hypersetup{colorlinks,linkcolor=red,citecolor=blue} 

\usepackage{dsfont,empheq}
\usepackage{multicol,multirow}

\usepackage{cite}

\usetikzlibrary{plotmarks,matrix,chains,scopes,fit,calc,shapes,positioning,decorations,intersections,arrows,backgrounds,shadows}

\newlength\FigureHeight
\newlength\FigureWidth
\pgfkeys{/pgf/number format/.cd,1000 sep={}}
\pgfplotsset{every axis legend/.append style={legend cell align=left,at={(0.02,0.97)},anchor=north west}}
\pgfplotsset{every axis plot/.append style={line width=1.5pt}} 
\pgfplotsset{every axis plot/.append style={mark options={solid,fill=white!80!.,line width=0.5pt},mark size = 3pt}} 

\definecolor{myDarkGreen}{rgb}{0.00000,0.58824,0.00000}%
\definecolor{uniform}{rgb}{0.00000,0.58824,0.00000}%
\definecolor{matched}{rgb}{1,0.58824,0.00000}%
\definecolor{mismatched}{rgb}{0.00000,0.44700,0.74100}%
\definecolor{AWGNreference}{rgb}{0.00000,0.58824,0.00000}%
\definecolor{XPMmodel}{rgb}{0,0,0}%
\definecolor{sims}{rgb}{0.00000,0.44700,0.74100}%

\pgfplotsset{AWGN_capacity/.style={color=gray,dotted}}
\pgfplotsset{16QAM_uniform/.style={color=blue,dashed}}
\pgfplotsset{16QAM_shaped/.style={color=blue,solid}}
\pgfplotsset{64QAM_uniform/.style={color=red,dashed}}
\pgfplotsset{64QAM_shaped/.style={color=red,solid}}
\pgfplotsset{256QAM_uniform/.style={color=brown,dashed}}
\pgfplotsset{256QAM_shaped/.style={color=brown,solid}}

\pgfplotsset{AWGNreference/.style={color=AWGNreference,dotted}}
\pgfplotsset{XPMmodel/.style={color=XPMmodel,dashed}}
\pgfplotsset{sims/.style={color=sims,only marks,mark=*,mark options={solid,fill=white}}}
\DeclareMathOperator*{\argmax}{argmax}

\newcounter{lemma}
\newtheorem{theorem}[lemma]{Theorem}
\newtheorem{corollary}[lemma]{Corollary}
\newtheorem{lemma}{Lemma}

\newtheorem{exampleplain}{Example}
\newtheorem{remark}{Remark}
\newenvironment{example}{\begin{exampleplain}}{~\hfill$\vartriangle$\end{exampleplain}}

\newcommand{\cd}{\cdot}
\newcommand{\mcIkb}{\mathcal{I}_{k}^{b}}
\newcommand{\dij}[0]{d_{ij}}\newcommand{\bdij}[0]{\boldsymbol{d}_{ij}}

\newcommand{\dip}[0]{d_{ip}}\newcommand{\bdip}[0]{\boldsymbol{d}_{ip}}

\newcommand{\set}[1]{\{#1\}}
\newcommand{\figref}[1]{Fig.~\ref{#1}}
\newcommand{\secref}[1]{Sec.~\ref{#1}}

\newcommand{\tr}[1]{\mathrm{#1}}
\newcommand{\ld}{\ldots}
\newcommand{\ms}[1]{\mathds{#1}}
\newcommand{\bxi}{\boldsymbol{\xi}}
\newcommand{\bc}{\boldsymbol{c}}
\newcommand{\bB}{\boldsymbol{B}}\newcommand{\bb}{\boldsymbol{b}}
\newcommand{\bX}{\boldsymbol{X}}\newcommand{\bx}{\boldsymbol{x}}
\newcommand{\bR}{\boldsymbol{R}}\newcommand{\br}{\boldsymbol{r}}
\newcommand{\bY}{\boldsymbol{Y}}\newcommand{\by}{\boldsymbol{y}}
\newcommand{\bZ}{\boldsymbol{Z}}\newcommand{\bz}{\boldsymbol{z}}

\newcommand{\bs}{\boldsymbol{s}}

\newcommand{\mcC}{\mathcal{C}}

\newcommand{\mcX}{\mathcal{X}}

\newcommand{\mcXkb}{\mathcal{X}_{k}^{b}}
\newcommand{\mcXko}{\mathcal{X}_{k}^{1}}
\newcommand{\mcXkz}{\mathcal{X}_{k}^{0}}
\newcommand{\Ns}{L}
\newcommand{\MCs}{D}
\newcommand{\GHs}{J}

\newcommand{\Rc}{R_\tr{c}}
\newcommand{\Ex}{\ms{E}}

\newcommand{\un}[1]{\underline{#1}}

\newcommand{\GMI}{\ensuremath{G}\xspace}
\newcommand{\MI}{\ensuremath{I}\xspace}

\newcommand{\tnr}[1]{{\textnormal{#1}}}

\pgfplotsset{compat=newest} 
\usepgfplotslibrary{external} 
\begin{document}
\title{Achievable Information Rates for Fiber Optics: Applications and Computations}
\IEEEspecialpapernotice{(Invited Paper)}

\author{Alex Alvarado,~\IEEEmembership{Senior Member,~IEEE,} Tobias~Fehenberger,~\IEEEmembership{Student Member,~IEEE}, Bin Chen,~\IEEEmembership{Member,~IEEE}, and Frans M. J. Willems,~\IEEEmembership{Fellow,~IEEE}. 
\thanks{A. Alvarado, B. Chen, and F. M. J. Willems are with the Signal Processing Systems (SPS) Group, Department of Electrical Engineering, Eindhoven University of Technology, 5600 MB Eindhoven, The Netherlands (e-mails: alex.alvarado@ieee.org, chen.bin.conan@gmail.com, f.m.j.willems@tue.nl)}
\thanks{Tobias Fehenberger is with the Institute for Communications Engineering, Technical University of Munich (TUM), 80333 Munich, Germany (\mbox{e-mail:}~tobias.fehenberger@tum.de}
\thanks{B. Chen is also with School of Computer and Information, Hefei University of Technology (HFUT), Hefei, China}
\thanks{This research has been supported in part by the Netherlands Organisation for Scientific Research (NWO) via the VIDI Grant ICONIC (project number 15685), by the Technical University of Munich (TUM) Graduate School Partnership Mobility Grant, and
by the National Natural Science Foundation of China (NSFC) under Grant 61701155.}
}

\maketitle

\begin{abstract}
In this paper, achievable information rates (AIR) for fiber optical communications are discussed. It is shown that AIRs such as the mutual information and generalized mutual information are good design metrics for coded optical systems. The theoretical predictions of AIRs are compared to the performance of modern codes including low-parity density check (LDPC) and polar codes. Two different computation methods for these AIRs are also discussed: Monte-Carlo integration and Gauss-Hermite quadrature. Closed-form ready-to-use approximations for such computations are provided for arbitrary constellations and the multidimensional AWGN channel. The computation of AIRs in optical experiments and simulations is also discussed.
\end{abstract}


\section{Introduction and Motivation}
\IEEEPARstart{T}{hrough} a series of revolutionary technological advances, optical transmission systems have supported the Internet's traffic growth for decades \cite{Richardson2010Science_CapacityCrunch}. Most of the large bandwidth available in fiber systems is in use \cite{Essiambre2012ProcIEEE_CapacityTrendsReview}, however, it seems like the capacity of the optical core network cannot keep up with the traffic growth \cite{Bayvel2016PhilTransRSocA_MaximizingCapacityReview,Winzer17JLT}. This makes information-theoretic analyses of optical fiber systems very important so as to maximize the information rates and spectral efficiencies of the nonlinear optical channel.

The optical fiber channel is effectively band-limited by the fiber loss profile and the operating range of the optical amplifiers \cite{Richardson2010Science_CapacityCrunch}. Because of this, the need of designing bandwidth-efficient transceivers is an active area of research. The most natural way of achieving this is via multi-level modulation combined with forward error correction (FEC), a combination known as coded modulation (CM). Different coded modulation \emph{flavors} exist, the most popular ones being trellis coded modulation \cite{Ungerboeck82}, bit-interleaved coded modulation (BICM) \cite{Zehavi92,Caire98,Fabregas08_Book,Alvarado15_Book,Alvarado2015_JLT}, and MLC \cite{Imai77,Wachsmann99}. FEC is usually either soft-decision (SD-FEC) or hard-decision (HD-FEC), which refers to the type of information passed to the FEC decoder: soft bits or hard bits, respectively.

\begin{figure}[tp]
\begin{center}
\includegraphics{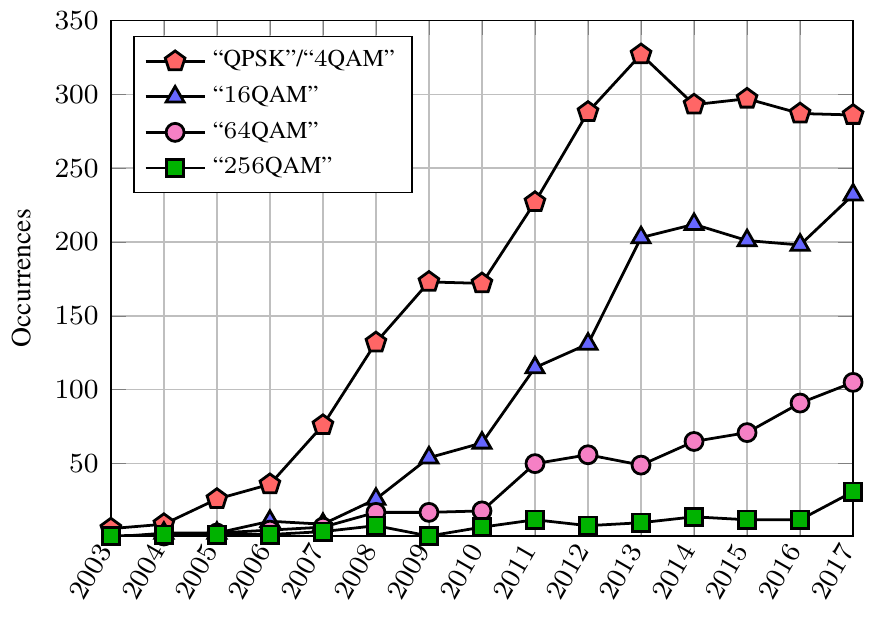}

\caption{Number of appearances of ``$M$QAM'' in OFC proceedings 2003--2017.}\label{OFC}
\end{center}
\end{figure}

CM has become a key technology for fiber-optical communications, as shown in Fig.~\ref{OFC}. This figure shows the number of occurrences of ``$M$QAM'' in the OFC proceedings in the last 15 years. Although the number of occurrences is a very rough metric (and can certainly be improved), this figure does show a clear trend: QPSK and 16QAM are here to stay, and 64QAM and 256QAM are slowly becoming more important.

Ungerboeck's celebrated trellis-coded modulation \cite{Ungerboeck82} was very popular because the receiver could find the most likely coded sequence using a single low-complexity decoder, that exploited the nonbinary (NB) trellis structure of the code. With the advent of powerful SD-FEC (such low-density parity-check (LDPC) codes \cite{Gallager63_Thesis}, turbo codes \cite{Berrou93}, and polar codes \cite{Arikan2009}), however, modern fiber optical CM transceivers use a receiver based on \emph{binary} FEC. When only one code is used, the scheme is known as BICM \cite{Zehavi92,Caire98,Fabregas08_Book,Alvarado15_Book,Alvarado2015_JLT}. When multiple codes are used, the system is called multilevel codes (MLC) \cite{Imai77,Wachsmann99}. In MLC, two decoding strategies are possible: Multi-stage decoding (MLC-MSD) \cite[Sec.~II]{Wachsmann99} and parallel, independent decoding of the individual levels (MLC-PDL) \cite[Sec.~VI-B]{Wachsmann99}. Both BICM and MLC-PDL are pragmatic (but suboptimal) approaches to CM where the detection is split into two stages: demap the bits independently, and then decode. 

Until a few years ago, pre-FEC bit-error rate (BER), symbol error rate (SER), Q-factor, and error vector magnitude (EVM) were the standard performance metrics in the optical communications community. Recently, however, this paradigm has been changing, and achievable information rates (AIRs) are becoming more popular. AIRs indicate the number of information bits per symbol that can be reliably transmitted through the channel and are at the core of Shannon's celebrated concept of \emph{channel capacity} \cite{Shannon48}. Because of the definition of AIRs, one of the key advantages of using them as performance metrics is that they are inherently related to FEC. Unlike uncoded metrics like the pre-FEC BER, SER, EVM, or Q-factor, AIRs give an indication of the amount of \emph{information bits} that can be reliably transferred through a channel. While uncoded metrics are related to bits before and after the demapper, AIRs deal with the information bits before and after FEC. 

In this paper, we discuss the use of AIRs as a tool to design CM transceivers with both NB and binary FEC. By means of simple examples, AIRs are shown to be very powerful design metrics. AIRs will be used to allow fair comparisons of different constellations, DSP, decoding and nonlinearity compensation techniques, and also for post-FEC error prediction.\footnote{AIRs can also be used to characterize optical transceivers, as done in \cite{Maher16}.} An AIR for CM based on NB codes or MLC-MSD is the mutual information (MI) \cite{Shannon48,Fehenberger15a,Liga16,Schmalen17}. For binary FEC (with BICM or with MLC-PDL), however, the relevant quantity is the GMI \cite{Caire98,Fabregas08_Book,Alvarado15_Book,Alvarado2015_JLT}. In this paper we consider equally likely symbols only, however, the extension of GMI to nonuniform symbols can be done, as shown in \cite[Sec.~VI]{Bocherer15}.

In this paper, we extend our results in \cite{Alvarado17OFCInvited} by also considering polar coded modulation with multilevel coding (MLC). We also give general ready-to-use expressions for approximating AIRs for multidimensional modulation formats. The paper is structured as follows. In \secref{Sec:General}, we discuss general aspects of AIRs. In \secref{Sec:IT}, we present an information-theoretic treatment of AIRs. In \secref{Sec:Computation}, computational aspects of AIRs are discussed.

\section{General Aspects of AIRs}\label{Sec:General}

Achievable information rates (AIRs) are information-theoretic quantities that are, by definition, linked to the amount of reliable information that can be transmitted through a given channel. This section gives examples of how AIRs can be used to predict key metrics of digital communication systems. At the end of the section, we highlight the underlying assumption of such an AIR analysis and the corresponding limitations.

\subsection{Three Applications of AIRs}\label{Sec:General:Applications}

In the following, applications of AIRs are are shown via three examples. The first use case is to predict the maximum throughput of an AWGN channel. Next, AIRs are investigated to compare the maximum reach for different modulation formats and DSP techniques in optical fiber transmission. In this context, a comparison of AIRs to LDPC and polar codes is performed. Finally, the accuracy of using MI and GMI for predicting the post-FEC BER is presented. In all cases, results with FEC encoding and decoding are included to show that the throughput of these coded systems follows the AIR predictions.

\begin{example}[Throughput Prediction]\label{Example.1}
Fig.~\ref{ThroughputPolar} shows the MI (solid green line) and GMI (dashed red line) for $256$QAM as well as the capacity of the 2D complex AWGN channel\footnote{We consider here 2D complex constellation formed as the Cartesian product of two $M$QAM constellation.}  (solid black line) as a function of SNR, which is defined in \secref{Sec:Prel.AWGN}. The curves in Fig.~\ref{ThroughputPolar} show that MI and GMI are very close to each other at high SNRs, whereas a clear gap is visible in the low- and medium-SNR range. This is the case for all square QAM constellations labeled by the binary reflected Gray code (BRGC) \cite{Gray1953Patent_BRGC}. For a particular SNR, the value of MI (and GMI) gives the maximum number of information bits that can be transmitted with a vanishing probability of error. 

AIRs are asymptotic metrics in the sense that they assume an ideal FEC with infinitely long codewords (see \secref{ssec:limitations} for details). Here we compare them to the performance of polar codes (lines with markers in Fig.~\ref{ThroughputPolar}) with a finite block length of $2^{12}=4096$~code bits. These polar codes allow a flexible adjustment of the code rate by freezing certain bit channels, i.e., any code rate can be achieved. Polar codes were used to implement two CM approaches: BICM and MLC-MSD \cite{Seidl2013_TCOM,Bocherer2017}.  In both cases the targeted code rates are the same: $\Rc\in\set{1/3,2/5,1/2,3/5,3/4,5/6,9/10}$. 

For BICM, a single binary polar code is used for encoding, and thus, the code rates under consideration are the same as the targeted rates. For the design of MLC-MSD with polar codes, however, multiple encoders and decoders need to be designed. This was done by matching the rates of each polar encoder to the conditional MIs such that the overall rate corresponds to the targeted code rate. More details about this are given in Sec.~\ref{MI}.

In BICM, the decoder is based on a single polar decoder. In MLC-MSD, each sub-decoder uses a standard successive cancellation decoding algorithm \cite[Sec.~VIII]{Arikan2009} to compute reliability information for each stage. Then, the soft information is passed from one sub-decoder to the next one. This process continues until all codes have been decoded. Note that MSD is similar to successive cancellation of each polar code in the sense that both decoding processes are implemented in a successive manner.

Fig.~\ref{ThroughputPolar} shows the required SNR for these polar-coded CM schemes to achieve a post-FEC bit error probability below $10^{-4}$ (markers). We observe that the MI can be used to predict the throughput of polar codes with MLC-MSD and that GMI predicts the performance of polar codes with BICM. The gap between AIRs and polar code results is due to the suboptimality (i.e., finite length) of the codes and can be decreased by increasing the code length or by performing list decoding \cite{Tal2015}.

\begin{figure}[t]
\centering
\includegraphics{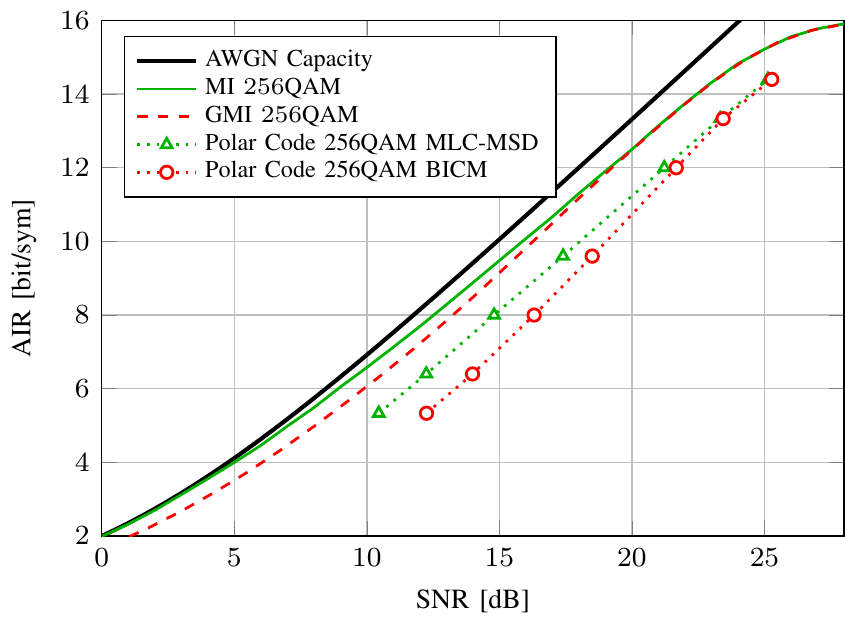}
\caption{MI (solid) and GMI (dashed) vs. SNR for the 2D complex AWGN channel and $256$QAM (maximum spectral efficiency of  $16$~bit/sym). The capacity of the AWGN channel in \eqref{C.MD.AWGN} is also shown (thick solid line). The results obtained using polar codes are shown with markers.}
\label{ThroughputPolar}
\end{figure}

\begin{figure}[t]
\centering
\includegraphics{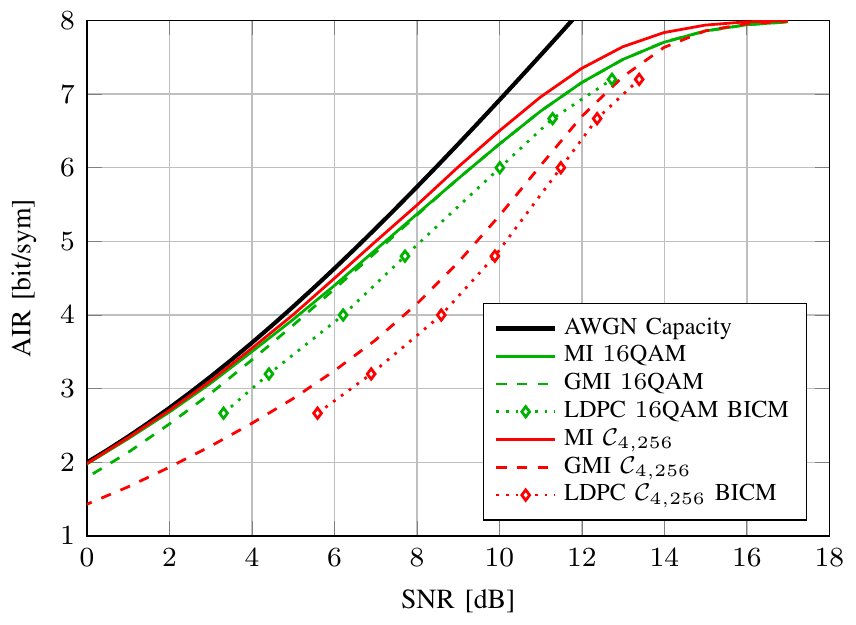}
\caption{MI (solid) and GMI (dashed) vs. SNR for the 2D complex AWGN channel. Two constellations with a maximum spectral efficiency of  $8$~bit/sym are considered: $16$QAM and $\mathcal{C}_{4,256}$. The capacity of the AWGN channel in \eqref{C.MD.AWGN} is also shown (thick solid line). The results obtained using LDPC codes from \cite{ETSI_EN_302_307_v121} are shown with markers.}
\label{ThroughputLDPC}
\end{figure}

Fig.~\ref{ThroughputLDPC} shows MI, GMI, and the results of binary LDPC codes with the same seven code rates used for polar codes in Fig.~\ref{ThroughputPolar}.\footnote{Each transmitted frame consists of $64800$ code bits. The decoder uses the message passing algorithm with $50$ iterations.} Note that the block lengths of the LDPC and polar coding schemes are chosen as significantly different to facilitate their design process. A detailed comparison of finite-length and complexity aspects of these FEC schemes is outside the scope of this paper. The two considered modulation formats are polarization-multiplexed $16$QAM as well as $\mathcal{C}_{4,256}$ \cite[Table~IV]{Welti74}, \cite[Table~I]{Conway83}, which is one of the most power-efficient 4D formats at asymptotically high SNRs \cite[Fig.~1~(a)]{Karlsson12OFC}. We observe that the MI of this optimization modulation format is larger than that of $16$QAM. This MI gain, however, does not translate into rate gains with LDPC codes because the GMI of $\mathcal{C}_{4,256}$ is lower than the one of $16$QAM. This figure suggests that binary LDPC codes follow the GMI prediction rather than the MI.
\end{example}

\begin{example}[Reach Increase Prediction]\label{Example.2}
Fig.~\ref{Reach} shows AIRs versus transmission distance and highlights the fact that AIRs can be used to predict the reach increase in optical systems, as demonstrated in \cite{Fehenberger15a} and \cite{Liga16}. For the optical system in \cite{Liga16}, Fig.~\ref{Reach} shows GMIs for two different modulation formats and two nonlinearity compensation techniques: electronic dispersion compensation (EDC) and ideal single-channel digital back propagation (DBP). These formats and compensation techniques can be directly compared to each other (for the same target AIR). These results show that, for multi-span multi-channel optical transmission, single-channel DBP with $64$QAM offers a potential reach increase of approximately $1100$~km at $8$~bit/sym with respect to EDC. Fig.~\ref{Reach} also shows the AIRs for polar and LDPC codes of various rates. Similar to Figs.~\ref{ThroughputPolar} and \ref{ThroughputLDPC}, these codes follow the trend of the GMI curves. The larger gap between polar codes and the GMI (larger than for LDPC codes) is mainly due to the difference in block lengths used: 4096 bits for polar codes and 64800 for LDPC codes. 

The results in Fig.~\ref{Reach} highlight the importance of using a strong FEC for multi-span systems.\footnote{Note that within the scope of these simulations, potentially different implementation penalties of 16QAM and 64QAM are not considered.} This figure shows that an SNR penalty of about 2~dB for polar codes (see Fig.~\ref{ThroughputPolar}) results in a large reach loss. Similar conclusions were obtained when comparing HD-FEC and SD-FEC in the context of adaptive optical networks in \cite{Alvarado2016Optnet}. This same approach was used in \cite{Fehenberger15a} to compare probabilistic shaping and different DSP techniques where is was shown for the first time that the distance increase by probabilistically-shaped QAM input and EDC are equivalent to those offered by DBP with uniform QAM. AIRs can therefore be used to guide the CM design, and also to decide on how to trade complexity and performance.

\begin{figure}[t]
\centering
  \includegraphics{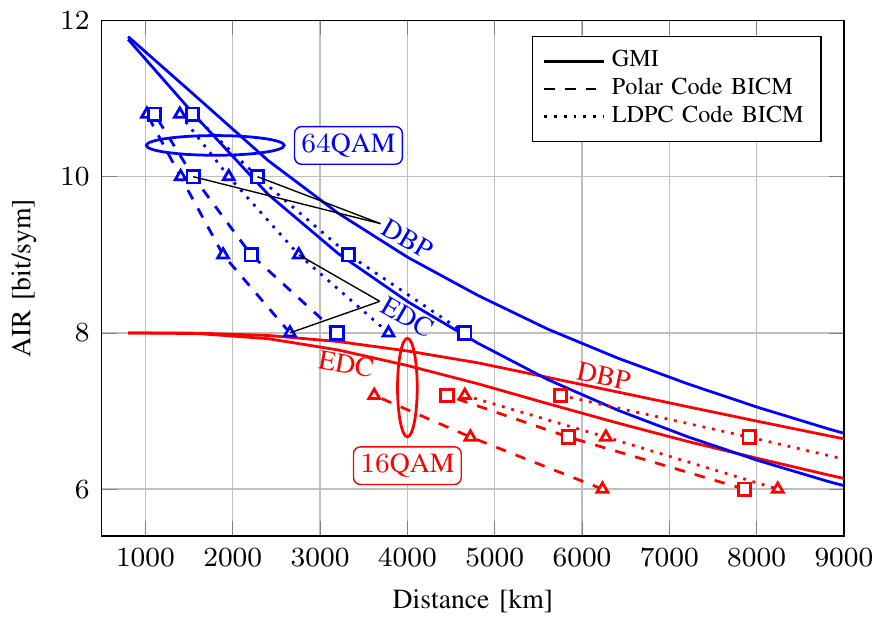}
\caption{GMI vs. transmission distance for optical fiber simulations of 5 densely spaced WDM channels with a per-channel symbol rate of 32~GBaud over multiple spans of 80~km single-mode fiber followed by an EDFA (see \cite[Table~1]{Liga16} for details). The markers show the reach achieved by polar and LDPC codes with rates $\Rc\in\set{2/3,3/4,5/6,9/10}$.}
\label{Reach}
\end{figure}
\end{example}

\begin{example}[Post-FEC Performance Prediction]\label{Example.3}
AIRs can also be used as decoding thresholds. The error probability after FEC can often be accurately predicted by considering the MI and GMI. Consider for example the complex AWGN channel, the NB-LDPC codes from \cite{Schmalen17}, and three different $8$QAM constellations from \cite[Fig.~3]{Schmalen17}. The post-FEC SER results for this NB-CM scheme are shown in Fig.~\ref{predictionNB}, where different markers represent different modulation formats and the asymptotic decoding thresholds are shown as dashed vertical lines. We observe from this figure that the MI (normalized by the number of bits per symbol $m=\log_2{M}$) is a good predictor of the post-FEC SER.\footnote{MI as a post-FEC BER predictor in optical communications can be traced to \cite{Leven2011PTL_softFECperf}. This idea was however proposed in \cite{Brueninghaus05,Wan06} in the context of wireless communications.} We conjecture that similar results can be obtained if MLC-MSD polar codes are considered. 

\begin{figure}[t]
\centering
  \includegraphics{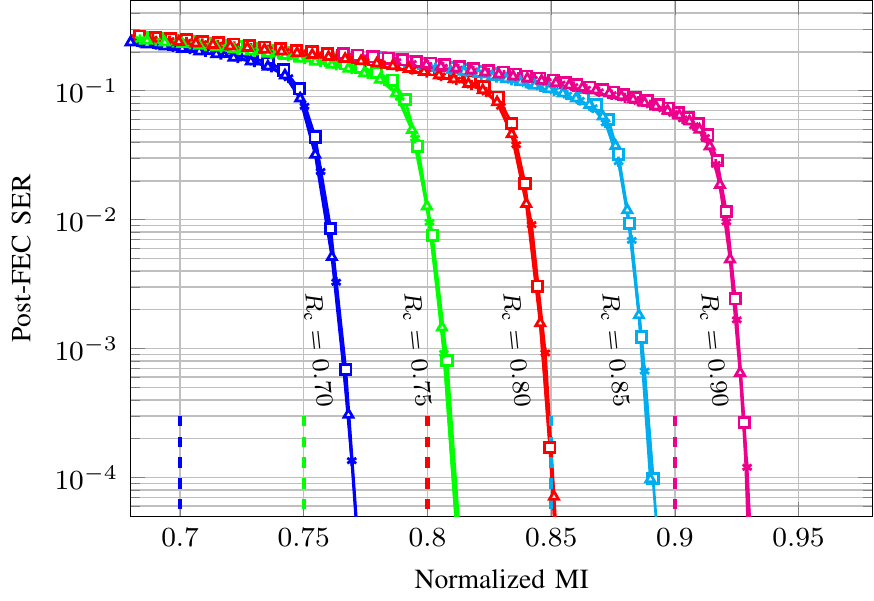}
\caption{Post-FEC prediction for NB from \cite{Schmalen17}. The vertical lines represent the normalized AIR for the corresponding code rates.}
\label{predictionNB}
\end{figure}

Fig.~\ref{predictionBinary} shows results for binary LDPC codes from the DVB-S2 standard \cite{ETSI_EN_302_307_v121}, where random puncturing was applied to obtained the rates shown. Three different modulation formats were considered: QPSK, $16$QAM, and $64$QAM (different markers). In this case the decoder is binary and the normalized GMI is the quantity that correctly predicts the post-FEC BER for all modulation formats. 

The main conclusion from Figs.~\ref{predictionNB} and \ref{predictionBinary} is that normalized MI and GMI are very good decoding thresholds for SD-FEC. Limitations to this method are outlined below in \secref{ssec:limitations}. For the considered binary LDPC codes, Table~\ref{tab:gmisummary} shows the overall code rate, which is the product of an outer staircase code with 6.25\% overhead (rate $1/1.0625=94.1\%$) \cite[Table~1]{ZhangKschischang2014JLT_Staircase} and the rate $\Rc$ of the LDPC code. For such a concatenated coding scheme, interleaving over different codewords is assumed to break up potential burst errors. The GMI is estimated at the HD-FEC decoding threshold of $\text{BER}=4.7\times10^{-3}$, which makes Table~\ref{tab:gmisummary} a ready-to-use lookup table to find the decoding threshold of the concatenation of inner DVB-S2 LDPC code and outer staircase code. The coding gaps in Table~\ref{tab:gmisummary} reflect the potential improvements from using stronger SD and HD FEC.

\begin{figure}[t]
\centering
  \includegraphics{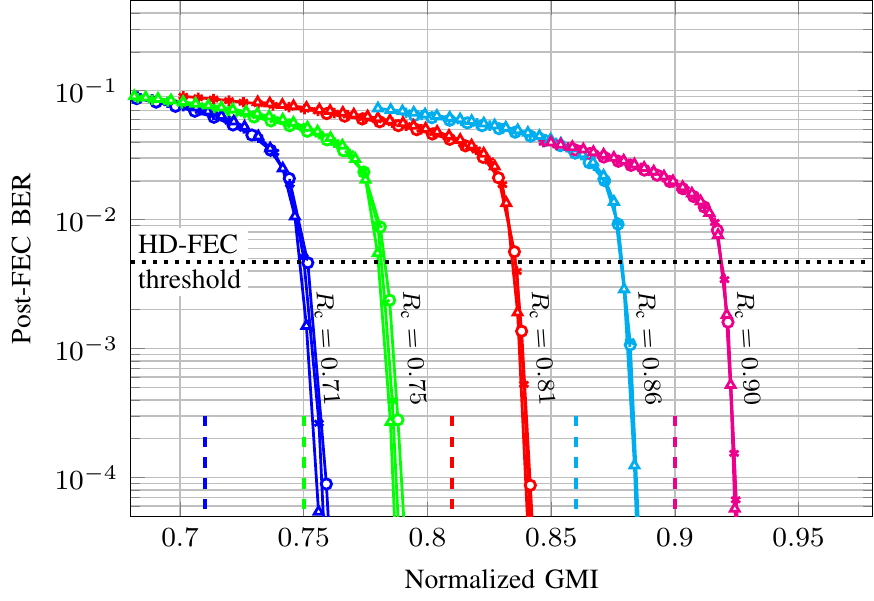}
\caption{Post-FEC prediction for binary LDPC codes. Some of the rates are achieved by randomly puncturing the codes defined in \cite{ETSI_EN_302_307_v121}. The vertical lines represent the normalized AIR for the corresponding code rates.}
\label{predictionBinary}
\end{figure}

\begin{table}[t]
\caption{Overall rate and normalized GMI required for LDPC codes with different code rates $\Rc$ that are concatenated with a staircase code to achieve a BER of $10^{-15}$ after decoding.}
\centering
\renewcommand{\arraystretch}{1.1}
\begin{tabular}{ c c c c}
\hline
  $\Rc$ & Overall Rate & GMI LDPC & Coding gap at threshold\\
\hline
${0.71}$ & $0.67$ & $0.75$ & $0.08$\\
${0.75}$ & $0.71$ & $0.78$  & $0.07$\\
${0.81}$ & $0.76$ & $0.84$ & $0.07$\\
${0.86}$ & $0.81$&$0.88$ & $0.07$\\
${0.9} $ & $0.85$&${0.92}$ & $0.07$\\
\hline
  \end{tabular}
  \label{tab:gmisummary}
\end{table}
\end{example}

\subsection{Limitations}\label{ssec:limitations}

Some important differences of AIRs and realistic FEC schemes must be considered that can limit the applicability of AIRs as performance predictors and decoding thresholds. For example, AIRs such as MI and GMI inherently assume a capacity-achieving FEC code with ideal maximum-likelihood (ML) decoding for which the complexity grows exponentially with the block length of the block code \cite[Sec.~5-1]{Ryan2009Book_ChannelCodes}. Decoding of almost all modern codes is performed with suboptimal low-complexity decoding algorithms (such as the sum-product algorithm \cite{Kschischang2001SP} for LDPC codes).\footnote{Note that successive cancellation decoding for polar codes  is ideal in the sense of achieving capacity. However, it does not perform well for short and moderate block lengths. The coding gap can be decreased by increasing the block length or by performing list decoding.}
When MI or GMI is used as decoding threshold for a FEC with a suboptimal decoding algorithm, trapping sets that result in an error floor must be treated with care. Such an error floor is the result of the considered code and decoding algorithm, and an AIR analysis with MI and GMI as it is carried out in this paper does not include this effect. Lastly, MI and GMI can be used as decoding thresholds only when the encoder and decoder pair in question is universal. This has been discussed in detail in \cite[Secs.~II-C and V]{Schmalen17}.

A significant performance difference of polar and LDPC codes to the AIRs exists because MI and GMI are asymptotic limits for infinitely long codewords. As this requirement is clearly impossible to fulfill for any practical FEC, the coding gap between AIR and FEC performance is increased. Depending on the application, it might be sometimes more insightful to include finite-length constraints in the analysis, for example via density evolution or via finite block length regime bounds (see for example \cite{polyanskiy2010channel}).


Another inherent assumption of the concatenated FEC schemes in this paper is that ideal interleavers are included between the inner and outer FEC. These interleaving blocks remove any error bursts that occur during transmission, which is a crucial assumption for applying the HD-FEC threshold as it is done in this paper. Furthermore, the results presented in Fig.~\ref{predictionBinary} (as well as those in \cite{Alvarado2015JLT_Replacing}) were generated using a random interleaver, i.e., a random interleaving permutation was generated for every transmitted codeword. This is in contrast to the results in \cite{Alvarado15a}, where no interleaver was used (the code bits from the LDPC code were cyclically mapped to the modulation format). The difference between these two approaches is that the former (the one used here) is practically less relevant, however, the obtained prediction is more accurate. These results should therefore be understood as the performance prediction for the code ensemble generated by the random realizations of the interleaver. On the other hand, by not using an interleaver (as done in \cite{Alvarado15a}), the post-FEC BER prediction is slightly worse for low code rates, but the results are highly practical. We conjecture that a very good trade-off between practical relevance and error prediction can be obtained if one single interleaver permutation is used all the time, as long as that permutation is randomly generated.\footnote{Another (less practical) alternative would be to use one interleaver that covers multiple LDPC codewords.}

As discussed in \secref{GMI}, if the LLRs are \emph{mismatched} to the communication channel or computed in an approximate manner, for example with the max-log approximation \cite{Viterbi1998JSAC_MaxLogApproximation}, lower AIRs can be obtained. This leads to the idea of correcting LLRs, which can also be connected to improving the performance of the decoder. For more details on this, we refer the reader to \cite[Ch.~7]{Alvarado15_Book} and references therein.

\section{Information-Theoretic Elements and AIRs}\label{Sec:IT}

\begin{figure*}[t]
\centering
  \includegraphics{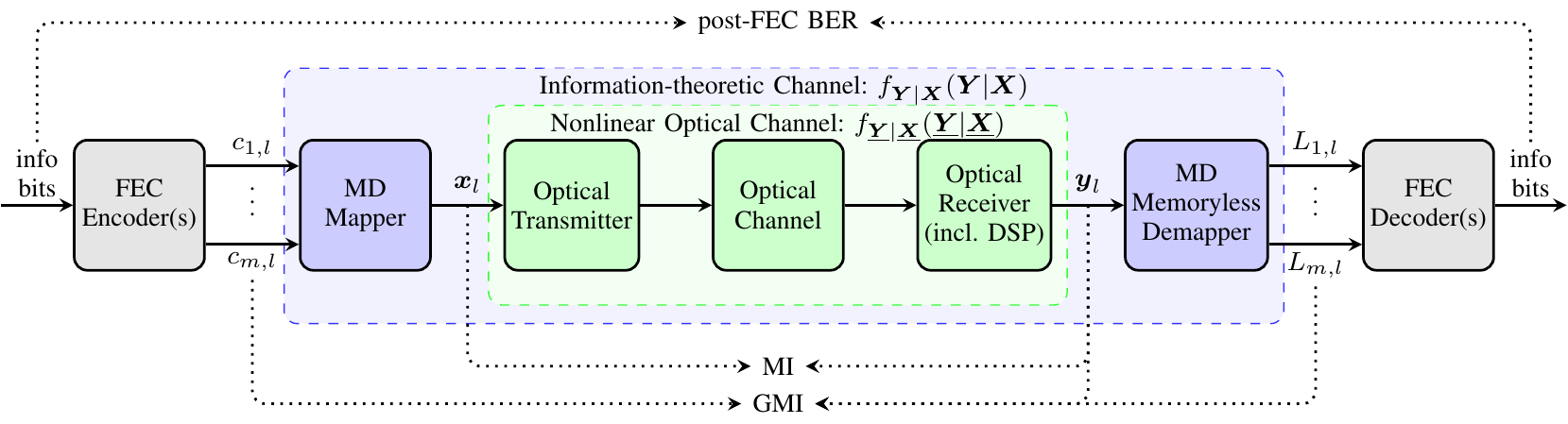}
    \caption{System model under consideration. The nonlinear optical channel is modeled using a channel with memory described by the PDF $f_{\un{\bY}|\un{\bX}}(\un{\bY}|\un{\bX})$ while the information-theoretic channel is modeled by the memoryless PDF $f_{\bY|\bX}(\bY|\bX)$. The MI and GMIs defined in \eqref{MI.def} and GMI in \eqref{GMI.def.General} respectively, are also shown. The GMI is also schematically shown between the code bits and LLRs, which is discussed in \secref{LLRGMI}.}
    \label{channel_blockdiagram}
\end{figure*}

\subsection{Coded Modulation}\label{ssec:it:cm}

Coded modulation (CM) comes in different \emph{flavors}, the most popular ones being TCM, NB-CM, MLC, and BICM. CM for optical communications is nowadays a well-established technique for spectrally efficient transmission. In this section, we focus on MLC and BICM due to their practical relevance. The FEC decoder used in CM also comes in two flavors: hard-decision (HD) and soft-decision (SD) decoding. In this paper, and due to its importance in next-generation optical transceivers, we only consider SD-FEC. For a comparison of AIRs between HD- and SD-FEC, we refer the reader to \cite{Liga16} and references therein.\footnote{Note that the AIRs for \emph{nonbinary} HD FEC studied in \cite{Liga16} are not achieved by standard Hamming-distance-based nonbinary HD FEC decoders, but instead, by codes that use some soft-decision for decoding. This was recently shown in \cite{Sheikh17OFC}, which was initially correctly pointed out in \cite[Footnote 10]{Liga16}. For more explanations on this, we refer the reader to \cite[Sec.~3.4]{Alvarado15_Book}.}

In general, the optical channel suffer from intersymbol and interpolarization interference (see inner part of \figref{channel_blockdiagram}). Most (if not all) optical receivers compensate for these effects, however, residual intersymbol interference (due to, e.g., Kerr nonlinearities after dispersion compensation) are typically neglected\footnote{When taking into account this residual interference, improved AIRs for fiber optics have been studied in, e.g.,\cite{Dar2014OptLetters_NewBounds,Fehenberger2015ECOC_ACF-XPM-CPE}}. Furthermore, each polarization is detected independently, and the soft information on the coded bits is calculated ignoring residual correlation between symbols in time. This is highlighted in \figref{channel_blockdiagram}, where a memoryless demapper is included at the receiver. The resulting ``information-theoretic channel'' is also shown in this figure, which can be interpreted as an ``average'' channel. This channel is fully characterized by the conditional PDF $f_{\bY|\bX}(\bY|\bX)$, where $\bX$ and $\bY$ are the transmitted and received symbols, respectively. We will return to these memoryless assumptions in \secref{Sec:Experiments}.

The transmitted symbols  $\bX$ are assumed to be multidimensional (MD) symbols with $N$ complex dimensions (or equivalently, with $2N$ real dimensions) drawn uniformly from a discrete constellation $\mcX$ with cardinality $M=2^m=|\mcX|$. The noisy received symbols are MD symbols with $N$ complex dimensions, i.e., $\bY\in\mathds{C}^N$. We consider constellation points $\bx_i=[x_{i,1},x_{i,2},\ldots,x_{i,N}]\in \mathds{C}^N$ with $i=1,2,\ld,M$. The difference between two symbols is defined as the vector $\bdij\triangleq\bx_i-\bx_j$, and the uniform input PMF by $P_{\bX}(\bx)=1/M$.\footnote{Notation: Random variables are denoted by capital letters $X$ and random vectors by boldface letters $\bX$. Their corresponding realizations are denoted by $x$ and $\bx$. The inner product between two vectors is denoted by $\left<\bx_1,\bx_2\right>$. Expectations are denoted by $\Ex[\cd]$ and the squared Euclidean norm is defined as $\|\bX\|^2=|X_{1}|^2+|X_{2}|^2+\ldots+|X_{N}|^2$. Conditional probability density functions (PDFs) are denoted by $f_{\bY|\bX}(\by|\bx)$. The imaginary unit is $\jmath\triangleq \sqrt{-1}$.} The most popular case in fiber optical communications is $N=2$, which corresponds to coherent communications using two polarizations of the light (4 real dimensions). 

Throughout this paper we consider four different CM structures based on SD-FEC. These four alternatives are shown in Table~\ref{table_CM}. The first one is NB-CM where a NB FEC is concatenated with a nonbinary modulation format (see Fig.~\ref{predictionNB}) and the second one is MLC-MSD (see Fig.~\ref{ThroughputPolar}). An AIR for these two cases is the MI. The third and fourth alternatives are BICM (see Fig.~\ref{ThroughputLDPC}) and MLC-PDL, for which the GMI is an AIR.

\begin{table*}[tb]
 \centering 
 \caption{Four different coded modulation \emph{flavors} based on SD FEC.}
 \label{table_CM}
 \renewcommand{\arraystretch}{1.2}
\begin{tabular}{cccc}
    \hline
    
    \hline
CM &  More Details     & AIR 	& Mathematical Expression     \\
    \hline
    
    \hline
NB FEC with NB modulation 	& \cite{Rong2008, Schmalen17,Beygi14} & MI & $I(\bX;\bY)$  \\
Binary FEC with MLC-MSD 	& \cite{Imai77}, \cite[Sec.~II]{Wachsmann99} 	& MI & $\sum_{k=1}^{m} I(B_k;\bY|B_1,\ld,B_{k-1})$ \\
\hline
Binary FEC with MLC-PDL 	& \cite[Sec.~VI-B]{Wachsmann99}			& GMI & $\sum_{k=1}^{m} I(B_k;\bY)$\\
Binary FEC with BICM 		& \cite{Zehavi92,Caire98,Fabregas08_Book,Alvarado15_Book} 						& GMI & $\sum_{k=1}^{m} I(B_k;\bY)$\\
    \hline
    
    \hline
\end{tabular}
\end{table*}

\subsection{Channel Capacity}\label{Capacity}

A \emph{coding scheme} consists of a codebook, an encoder, and a decoder. The codebook is the set of codewords that can be transmitted through the channel, where each codeword is a sequence of symbols. The encoder is a one-to-one mapping between the information sequences and codewords. The decoder is a deterministic rule that maps the noisy channel observations onto an information sequence.

A rate, in bits per MD symbol (or simply bit/sym), is said to be \emph{achievable} at a given block length $L$ and for a given average error probability $\varepsilon$ if there exists a coding scheme whose average error probability is below $\varepsilon$. The \emph{channel capacity} introduced by C. E. Shannon in 1948 \cite{Shannon48} is the largest AIR for which a coding scheme with vanishing error probability exists, in the limit of large block length. In simple words, the channel capacity represents the number of bits that can be reliably ``pushed'' through a given channel. Shannon also proved that the channel capacity cannot be exceeded, i.e., rates above the channel capacity cannot be reliable transmitted.\footnote{This is known in the information theory literature as the converse of the channel coding theorem which was proven in its strong form in \cite{Wolfowitz1957coding}.} For a more detailed and precise description of these concepts in the context of optical communications, we refer the reader to \cite{Agrell16RS}. In what follows, we discuss two AIRs: MI and GMI.

\subsection{Mutual Information}\label{MI}

Let $\mcC$ be the binary codebook used for transmission and $\un{\bc}$ denote the transmitted codewords as
\begin{align}
\un{\bc} = 
\left[
\begin{matrix}
c_{1,1} & c_{1,2} & \ld & c_{1,\Ns}\\
\vdots	&\vdots	& \ddots & \vdots \\
c_{m,1} & c_{m,2} & \ld & c_{m,\Ns}\\
\end{matrix}
\right],
\end{align}
where the block length (in symbols) is $L$.

Furthermore, let $\bB=[B_{1},\ld,B_{m}]$ be a random vector representing the transmitted bits $[c_{1,l},\ld,c_{m,l}]$ at any time instant $l$, which are mapped to the corresponding symbol $X_{l}\in\mcX$ with $l=1,2,\ld,\Ns$. Assuming a memoryless channel, the optimal symbol-wise (SW) ML receiver chooses the transmitted codeword based on an observed sequence $\un{\by}=[\by_{1},\ld,\by_{L}]$ according to the rule
\begin{align}\label{rule.ML}
\un{\bc}^{\tnr{sw}} \triangleq \argmax_{\un{\bc}\in\mcC}\sum_{l=1}^{\Ns}\log f_{\bY|\bB}(\by_{l}|c_{1,l},\ld,c_{m,l})
\end{align}
where $f_{\bY|\bB}(\by|\bb)=f_{\bY|\bX}(\by|\bx)$ is the channel law.

Shannon's channel coding theorem states that reliable transmission with the SW decoder in \eqref{rule.ML} is possible at arbitrarily low error probability if the combined rate of the binary encoder and mapper (in information bit per MD symbol) is below the MI $I(\bX;\bY)$, i.e., if $\Rc m\leq I(\bX;\bY)$. In other words, for any memoryless channel, the largest achievable rate is the MI defined as
\begin{align}\label{MI.def}
\MI = I(\bX;\bY) \triangleq \Ex_{\bX,\bY}\left[\log_{2}\frac{f_{\bY,\bX}(\bY,\bX)}{f_{\bY}(\bY)f_{\bX}(\bX)}\right]
\end{align}
where $\Ex_{\bX,\bY}$ denotes the expectation with respect to both $\bX$ and $\bY$.

The MI in \eqref{MI.def} for discrete constellations and equally likely symbols can be expressed as
\begin{align}
\label{MI.general.0}
\MI	& =\Ex_{\bX,\bY}\left[\log_2\frac{f_{\bY|\bX}(\bY|\bX)}{f_{\bY}(\bY)}\right] \\
\label{MI.general.1}
&=\frac{1}{M}\sum_{i=1}^{M} \int_{\mathds{C}^N}f_{\bY|\bX}(\by|\bx_i) \log_{2}\frac{f_{\bY|\bX}(\by|\bx_i)}{\frac{1}{M}\sum_{j=1}^{M}f_{\bY|\bX}(\by|\bx_j)}\, \tnr{d}\by
\end{align}
where the integral is an MD integral over the $N$-dimensional complex space, and \eqref{MI.general.1} follows from the law of total probability. The MI for an arbitrary MD memoryless channel can then be expressed as
\begin{align}
\label{MI.general.2}
\MI
&=m+\frac{1}{M}\sum_{i=1}^{M} \int_{\mathds{C}^N}f_{\bY|\bX}(\by|\bx_i) g_i^\MI(\by)\, \tnr{d}\by,
\end{align}
with
\begin{align}\label{MI.gi}
g_i^\MI(\by) = \log_{2}\frac{f_{\bY|\bX}(\by|\bx_i)}{\sum_{j=1}^{M}f_{\bY|\bX}(\by|\bx_j)}.
\end{align}
The MI is an AIR for NB-CM. It is also an AIR for MLC-MSD, which can be shown by the chain rule of mutual information
\begin{align}\label{ChainRule}
I(\bX;\bY)=\sum_{k=1}^{m} I(B_k;\bY|B_1,\ld,B_{k-1})
\end{align}
The code rates of the polar coded MLC-MSD in \secref{Sec:General:Applications} were designed to match the $m$ bit-wise conditional MIs in \eqref{ChainRule}. This is also shown in the second row of Table~\ref{table_CM}.

\subsection{Generalized Mutual Information}\label{GMI}

Bit-wise (BW decoders considered in this paper (BICM and MLC-PDL) split the decoding process. First, L-values are calculated, and then, one or multiple binary SD decoder are used. More precisely, the BW decoder rule is
\begin{align}\label{rule.BW}
\un{\bc}^{\tnr{bw}} \triangleq \argmax_{\un{\bc}\in\mcC}\sum_{l=1}^{\Ns}\log \prod_{k=1}^{m}f_{\bY|B_{k}}(\by_{l}|c_{k,l}).
\end{align}
The BW decoding rule in \eqref{rule.BW} is not the same as the ML rule in \eqref{rule.ML} and the MI is in general not an achievable rate with a BW decoder.\footnote{An exception is Gray-mapped QPSK with noise added in each quadrature independently. In this case, the detection can be decomposed into the detection of two BPSK constellations, and thus, SW and BW decoders are identical.}

The BW decoder can be cast into the framework of a mismatched decoder\footnote{Mismatched decoding theory\cite{Kaplan93,Merhav94,Ganti00} can be used to study AIRs when the detector is based on a suboptimal decoding rule. This technique has been used for example in \cite{Secondini2013JLT_AIR,Eriksson16,Schmalen17} in the context of optical communications.} by considering a symbol-wise metric\cite{Martinez08}
\begin{align}\label{metric.BW}
q(\bb,\by) \triangleq \prod_{k=1}^{m}f_{\bY|B_{k}}(\by|b_{k}).
\end{align}
Using this mismatched decoding formulation, the BW rule in \eqref{rule.BW} can be expressed as 
\begin{align}\label{rule.BW.2}
\un{\bc}^{\tnr{bw}} = \argmax_{\un{\bc}\in\mcC}\sum_{l=1}^{\Ns}\log q(\bb_{l},\by_{l})
\end{align}
where with a slight abuse of notation we use $\bb_{l}=[c_{1,l},\ld,c_{m,l}]^{T}$. Similarly, the ML decoder in \eqref{rule.ML} can be seen as a mismatched decoder with a metric $q(\bb_{l},\by_{l})=f_{\bY|\bB}(\by_{l}|\bb_{l})=f_{\bY|\bX}(\by_{l}|\bx_{l})$ which is ``matched'' to the channel. Using this interpretation, the BW decoder uses metrics matched to the bits $f_{\bY|B_{k}}(\by|b_{k})$, but not matched to the actual (symbol-wise) channel.

An achievable rate for a BW decoder is the GMI, defined as \cite[Eqs.~(59)--(60)]{Martinez09} \cite[(4.34)--(4.35)]{Alvarado15_Book}
\begin{align}\label{GMI.def.General}
\GMI & \triangleq \max_{s\geq 0} \Ex_{\bB,\bY}\left[ \log_{2}\frac{q(\bB,\bY)^{s}}{\sum_{\bb\in\set{0,1}^{m}}P_{\bB}(\bb)q(\bb,\bY)^{s}}\right].
\end{align}
The GMI expression in \eqref{GMI.def.General} is general in the sense that it holds for any metric $q(\bB,\bY)$ and for any symbol distribution $P_{\bB}(\bb)$. To simplify this expression, we make two assumptions. First, we assume that the bits $B_{1},\ld,B_{m}$ are independent, and second, that the receiver uses the BW metric in \eqref{metric.BW}. The first assumption is valid for any encoder that induces a uniform symbol distribution (which is the case we consider in this paper), and is valid for any constellation and labeling. The second assumption is valid for BICM and MLC-PDL where LLRs are computed for each bit independently (more details about this are given in Sec.~\ref{LLRGMI}).

Under these two assumptions, \eqref{GMI.def.General} can be expressed as \cite[Theorem~4.11]{Alvarado15_Book} and\cite[Corollary~4.12]{Alvarado15_Book}
\begin{align}
\label{GMI.BW.0}
\GMI & = \max_{s\geq 0} \sum_{k=1}^{m}\Ex_{B_{k},\bY}\left[ \log_{2}\frac{f_{\bY|B_{k}}(Y|B_{k})^{s}}{\frac{1}{2}\sum_{b\in\set{0,1}}f_{\bY|B_{k}}(\bY|b)^{s}}\right]\\
\label{GMI.BW.2}
		& = \sum_{k=1}^{m} I(B_{k};\bY),
\end{align}
where the optimization over $s$ in this case gives $s=1$.

The GMI in \eqref{GMI.BW.2} can be written as
\begin{align}
\GMI 
&=\sum_{k=1}^{m}\Ex_{B_k,\bY}\left[\log_2\frac{f_{\bY|B_{k}}(\bY|B)}{f_{\bY}(\bY)}\right]\\
\label{GMI.Integral.2}
		& = \frac{1}{M}\sum_{k=1}^{m}\sum_{b\in\set{0,1}}\sum_{i\in\mcIkb} \int_{\mathds{C}^N}f_{\bY|\bX}(\by|\bx_{i}) \cd \nonumber \\
		&\qquad\qquad\qquad\qquad \log_{2}\frac{\sum_{j\in\mcIkb}f_{\bY|\bX}(\by|\bx_{j})}{\frac{1}{2}\sum_{p=1}^{M}f_{\bY|\bX}(\by|\bx_{p})}\, \tnr{d}\by,
\end{align}
where $\mcIkb\subset\set{1,2,\ld,M}$ with $|\mcIkb|=M/2$ is the set of indices of constellation points whose binary label is $b$ at bit position $k$, and \eqref{GMI.Integral.2} follows by using the law of total probability. 

In analogy to \eqref{MI.general.2} and \eqref{MI.gi}, the GMI for an arbitrary MD memoryless channel can then be expressed as
\begin{align}
\label{GMI.Integral.3}
\GMI 	& = m+\frac{1}{M}\sum_{k=1}^{m}\sum_{b\in\set{0,1}}\sum_{i\in\mcIkb} \int_{\mathds{C}^N}f_{\bY|\bX}(\by|\bx_{i}) g_{i}^\GMI(\by)\, \tnr{d}\by,
\end{align}
with
\begin{align}\label{GMI.gi}
g_{i}^\GMI(\by)=\log_{2}\frac{\sum_{j\in\mcIkb}f_{\bY|\bX}(\by|\bx_{j})}{\sum_{p=1}^{M}f_{\bY|\bX}(\by|\bx_{p})},
\end{align}
The importance of the GMI in \eqref{GMI.Integral.3}--\eqref{GMI.gi} is that it represents an AIR for \emph{bit-wise receivers} such as BICM and MLC-PDL. In both cases, the decoding is suboptimal because the dependency between the bits in a symbol is ignored (see \eqref{metric.BW}). In view of \eqref{ChainRule}, the GMI in \eqref{GMI.BW.2} can be interpreted as a sum of unconditional bit-wise MIs. These MIs can be used to select the code rates in a MLC-PDL scheme, as shown in the third row of Table~\ref{table_CM}.

\begin{remark}
The GMI has not been proven to be the largest achievable rate for the BICM receiver. For example, a different achievable rate called the LM rate has been studied in \cite[Part~I]{Peng12_Thesis}. Finding the largest achievable rate with a BW decoder remains an open research problem. Despite this cautionary statement, the GMI is known to predict well the performance of CM transceivers based on capacity-approaching \mbox{SD-FEC} decoders, as shown in \secref{Sec:General:Applications}. 
\end{remark}

\subsection{MI and GMI for AWGN Channel}\label{Sec:Prel.AWGN}

We now consider the multidimensional memoryless AWGN channel $\bY=\bX+\bZ$, where $\bZ=[Z_{1},Z_{2},\ldots,Z_{N}]$ is a random vector whose entries are independent, complex, zero-mean, circularly symmetric Gaussian random variables. The total variance of the noise is therefore $\sigma_{\bz}^2=\Ex_{\bZ}[\|\bZ\|^2]$, where the square norm is defined as $\|\bZ\|^2=|Z_{1}|^2+|Z_{2}|^2+\ldots+|Z_{N}|^2$. The value of $\sigma_{\bz}^2$ is the total variance, and thus, the channel law is given by
\begin{align}\label{MD.AWGN.PDF}
f_{\bY|\bX}(\by|\bx) = \frac{1}{(\pi\sigma_{\bz}^2/N)^{N}}\exp{\left(-\frac{\|\by-\bx\|^2}{\sigma_{\bz}^2/N}\right)},
\end{align}
where $\sigma_{\bz}^2/N$ corresponds to the noise variance per complex dimension. 
The total transmitted power is given by $\sigma_{\bx}^2\triangleq\Ex_{\bX}[\|\bX\|^2]$, and thus, the SNR is defined as $\tnr{SNR}\triangleq\sigma_{\bx}^2/\sigma_{\bz}^2$. The channel capacity (in bit/sym) of the MD AWGN channel under an average power constraint is given by
\begin{align}\label{C.MD.AWGN}
C=N \log_2{\left(1+\tnr{SNR}\right)}.
\end{align}


The next two theorems show general expressions for the MI and GMI for the MD AWGN channel.
\begin{theorem}\label{MI.MDAWGN.Theo}
The MI for the MD AWGN channel is
\begin{align}
\nonumber
\label{MD.MI.general.5}
 \MI       & = m-\frac{1}{M}\sum_{i=1}^{M} \int_{\mathds{C}^N} f_{\bZ}(\bz) \cd \\
  	&\qquad \log_{2}{\sum_{j=1}^{M}\exp{\left(-\frac{\|\bdij\|^2+2\Re\set{\left<\bz,\bdij\right>}}{\sigma_{\bz}^2/N}\right)}} \, \tnr{d}\bz,
\end{align}
where $\bZ$ is a circularly symmetric zero-mean complex Gaussian random vector with (total) variance $\sigma_{\bz}^2$ and $\bdij \triangleq\bx_i-\bx_j$ as defined in Sec.~\ref{ssec:it:cm}.
\end{theorem}
\begin{IEEEproof}
See Appendix~\ref{Appendix.MI.MDAWGN.Theo}.
\end{IEEEproof}

\begin{theorem}\label{GMI.MDAWGN.Theo}
The GMI for the MD AWGN channel is
\begin{align}
\label{MD.GMI.general.5}
\nonumber
\GMI       & = m-\frac{1}{M}\sum_{k=1}^{m}\sum_{b\in\set{0,1}}\sum_{i\in\mcIkb} \int_{\mathds{C}^N} f_{\bZ}(\bz) \cd \\
&\qquad \log_{2}\frac{\sum_{p=1}^{M}\exp\left(-\frac{\|\bdip\|^{2}+2\Re\set{\left<\bz,\bdip\right>}}{\sigma_{\bz}^2/N}\right)}{\sum_{j\in\mcIkb}\exp\left(-\frac{\|\bdij\|^{2}+2\Re\set{\left<\bz,\bdij\right>}}{\sigma_{\bz}^2/N}\right)}\, \tnr{d}\bz.
\end{align}
\end{theorem}
\begin{IEEEproof}
See Appendix~\ref{Appendix.GMI.MDAWGN.Theo}.
\end{IEEEproof}

In general, $\MI\geq\GMI$ \cite[Theorem~4.24]{Alvarado15_Book}\footnote{The condition of i.i.d. bits in \cite[Theorem~4.24]{Alvarado15_Book} is not necessary---only independence is needed.}, where the rate penalty $\MI-\GMI$ can be understood as the penalty caused by the use of a suboptimal (BW) decoder. This rate penalty, however, is known to be small for Gray-labeled constellations \cite[Fig.~4]{Caire98}, \cite{Agrell10b,Alvarado12b}, \cite[Sec.~IV]{Alvarado11b}, as already discussed in Example~\ref{Example.1}.

\begin{figure}[t]
\centering
  \includegraphics{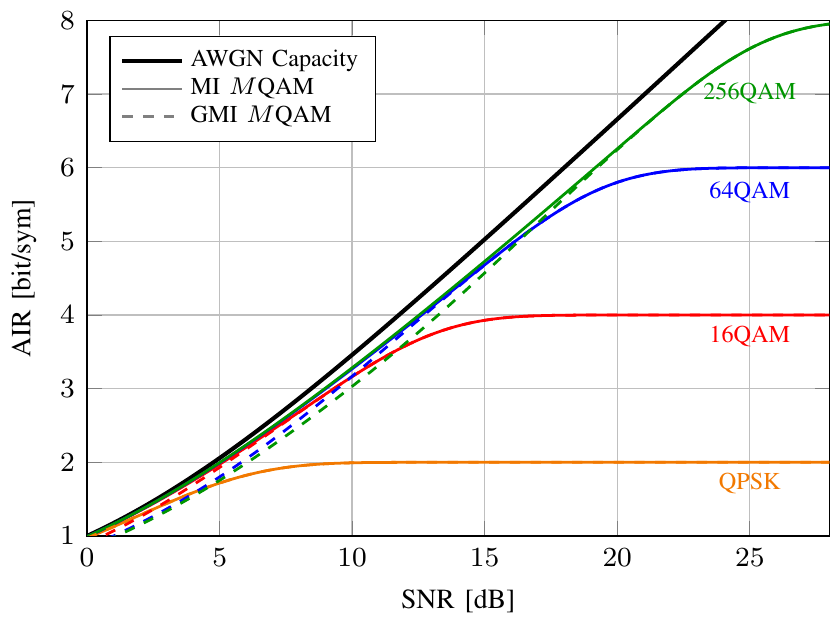}
\caption{MI (solid) and GMI (dashed) vs. SNR for the complex AWGN channel ($N=1$). The labeling rule for the GMI is the BRGC. The capacity of the AWGN channel in \eqref{C.MD.AWGN} is also shown (thick solid line).}
\label{MI_GMI_AWGN}
\end{figure}

\begin{example}[MI and GMI for QAM Constellations and AWGN]\label{Example.4}
\figref{MI_GMI_AWGN} shows MI and GMI for $M$QAM constellations over the complex AWGN channel. These AIRs were calculated using the numerical integration method described in \secref{Sec:GH}. When the BRGC is used as labeling rule, the GMI is very close to MI at high SNRs.\footnote{The asymptotic optimality of Gray codes for arbitrary real constellations has been recently proven in \cite[Theorem~11]{Alvarado12b}.} This means that the performance penalty from bit-wise processing (i.e., BICM and MLC-PDL) is very small, with a simultaneous significant reduction in complexity (for BICM). We further observe in \figref{MI_GMI_AWGN} that the GMI curves cross, i.e., there is an SNR range in which a QAM format with lower $M$ has a larger GMI than QAM of larger cardinality. This is due to the suboptimal (BW) demapping-decoding for the GMI and hence not present for the MI.
\end{example}

\subsection{LLR-based GMI}\label{LLRGMI}

The GMI was defined in \secref{GMI} in terms of the channel observations $\bY$. Most SD-FEC decoder used in BW receivers operate based on ``soft bits'', also known as L-values or simply log-likelihood ratios (LLRs). These soft bits are real numbers that represent the probability of the code bits. More precisely, the magnitude of an LLR indicates the reliability: an LLR close to zero means that no reliable information on that bit is available. On the other hand, a very large and positive (negative) LLR means that we are very certain the transmitted code bit was a one (zero). Because LLRs are simply another way of representing information, the GMI can be defined and analyzed in terms of these LLRs, as schematically shown in \figref{channel_blockdiagram}. We discuss precisely this in this section.

Assuming equally likely symbols, the LLRs for bit-wise decoding are at any time instant $l$ calculated as
\begin{align}\label{LLR.general}
L_{k}	& = \log\frac{\sum_{\bx\in\mcXko}f_{\bY|\bX}(\by|\bx)}{\sum_{\bx\in\mcXkz}f_{\bY|\bX}(\by|\bx)},
\end{align}
where $k=1,\ld,m$ and $\mcXkb\subset\mcX$ is the set of constellation symbols labeled by a bit $b\in\set{0,1}$ at bit position $k$.
To alleviate the computational complexity of \eqref{LLR.general}, the well-known max-log approximation is often used \cite{Viterbi98}
\begin{align}\label{LLR.max-log}
L_{k}	& \approx \log\frac{\max_{\bx\in\mcXko}f_{\bY|\bX}(\by|\bx)}{\max_{\bx\in\mcXkz}f_{\bY|\bX}(\by|\bx)}.
\end{align} 
In the case of AWGN channels and square QAM constellations, the max-log approximation results in piecewise linear relationships between the received symbol and the LLRs. This in turn makes its implementation very simple, which is partly why the max-log approximation is very popular in practice.

For the MD AWGN channel in \eqref{MD.AWGN.PDF}, the exact L-values in \eqref{LLR.general} and the max-log L-values in \eqref{LLR.max-log} are calculated, respectively, as
\begin{align}\label{LLR.sum-exp.AWGN}
L_{k}	& = \log\frac{\sum_{\bx\in\mcXko}\exp\left(-\frac{\|\by-\bx\|^{2}}{\sigma_{\bz}^2/N}\right)}{\sum_{\bx\in\mcXkz}\exp\left(-\frac{\|\by-\bx\|^{2}}{\sigma_{\bz}^2/N}\right)}
\end{align} 
and
\begin{align}
\label{LLR.max-log.AWGN}
L_{k}	& \approx \frac{\sigma_{\bz}^2}{N} \left(\min_{\bx\in\mcXkz}\|\by-\bx\|^{2}-\min_{\bx\in\mcXko}\|\by-\bx\|^{2}\right).
\end{align}

For equally likely symbols, and regardless of the LLR calculation and channel under consideration, the GMI in \eqref{GMI.BW.0} can be expressed as
\begin{align}
\label{GMI.general.LLR}
\nonumber
\GMI	& = m- \\
& \min_{s\geq 0}\sum_{k=1}^{m}\sum_{b\in\set{0,1}}\frac{1}{2}\int_{-\infty}^{\infty}f_{L_k|B_k}(l|b)\log_{2}\bigl(1+e^{s(-1)^{b}l}\bigr)\,\tnr{d}l.
\end{align}

\begin{remark}\label{Remark.1}
For the specific case when the L-values are calculated using \eqref{LLR.general}, $I(B_{k};Y)=I(B_{k};L_{k})$ \cite[Theorem~4.21]{Alvarado15_Book}, and thus, the GMI in \eqref{GMI.BW.2} becomes
\begin{align}\label{GMI.LLRs}
\GMI	& = \sum_{k=1}^{m} I(B_{k};L_{k})
\end{align}
i.e., the GMI is a sum of \emph{bit-wise} MIs between code bits and L-values.  The equality in \eqref{GMI.LLRs} does not hold, however, if the L-values were calculated using the max-log approximation \eqref{LLR.max-log}, or more generally, if the L-values were calculated using  any other approximation. For example, when max-log L-values are considered, it is possible to show that there is a loss in achievable rate. Under certain conditions, this loss can be recovered by correcting the max-log L-values, as shown in \cite{Jalden10,Nguyen11,Szczecinski12a}.
\end{remark}

LLRs mismatched to the channel or computed with some approximation result in a rate loss. Different LLR correction strategies can be used to improve the rate as well as the performance of the FEC decoder. Such a scenario was for example considered in \cite{Alvarado16ECOC}, where the LLRs for a channel subject to both additive and phase noise were matched only to the additive noise part of the channel. It was shown for this channel that different rates can be obtained depending on the type of correction used, and that the decoding performance can be improved by LLR scaling. For more details on this, we refer the reader to \cite{Alvarado16ECOC} and \cite[Ch.~7]{Alvarado15_Book} and references therein.

\section{Computation Methods for AIRs}\label{Sec:Computation}

The MI and GMI for the AWGN channel can be approximated using Monte-Carlo integration but also via Gauss-Hermite quadrature. These two methods are described in the following sections. The numerical computation of AIRs for the nonlinear optical fiber channel is discussed in \secref{Sec:Experiments}.

\subsection{Monte-Carlo Integration}\label{Sec:MC}

Monte-Carlo integration can be used to approximate an integral via a finite sum \cite[Chap.~9]{Tranter2004Book_MonteCarloPrinciples}, namely,
\begin{align}\label{MC}
\int_{\mathbb{C}^N} f_{\bR}(\br) g(\br) \,\tnr{d}\br \approx \frac{1}{\MCs} \sum_{n=1}^{\MCs} g(\br^{(n)}),
\end{align}
where $\bR$ is an arbitrary random vector defined over $\mathbb{C}^N$. In \eqref{MC}, $g(\br):\mathbb{C}^N\to\mathbb{R}$ is an arbitrary real-valued function and $\br^{(n)}$ with $n=1,\ld,\MCs$ are samples from the distribution $f_{\bR}(\br)$. In view of \eqref{MC}, Monte-Carlo approximations of \eqref{MD.MI.general.5} and \eqref{MD.GMI.general.5} are readily obtained, as shown in the following.

The following corollary gives Monte-Carlo approximations of the MI and GMI for the MD AWGN channel. The proof of this corollary follows directly from Theorems~\ref{MI.MDAWGN.Theo} and \ref{GMI.MDAWGN.Theo} together with \eqref{MC}.

\begin{corollary}[Monte-Carlo Approximations]\label{MD.MC.Theo}
The MI and GMI for the MD AWGN channel can be approximated as
\begin{align}\label{MD.MI.MC}
\nonumber
\MI & \approx m -\frac{1}{M}\sum_{i=1}^{M}\frac{1}{\MCs}\sum_{n=1}^{\MCs} \\
&\qquad \log_{2}{\sum_{j=1}^{M}\exp{\left(-\frac{\|\bdij\|^2+2\Re\set{\left<\bz^{(n)},\bdij\right>}}{\sigma_{\bz}^2/N}\right)}}\\
\label{MD.GMI.MC}
\nonumber
\GMI & \approx m-\frac{1}{M}\sum_{k=1}^{m}\sum_{b\in\set{0,1}}\sum_{i\in\mcIkb} \frac{1}{\MCs} \sum_{n=1}^{\MCs}\\
&\qquad
\log_{2}\frac{\sum_{p=1}^{M}\exp\left(-\frac{\|\bdip\|^{2}+2\Re\set{\left<\bz^{(n)},\bdip\right>}}{\sigma_{\bz}^2/N}\right)}{\sum_{j\in\mcIkb}\exp\left(-\frac{\|\bdij\|^{2}+2\Re\set{\left<\bz^{(n)},\bdij\right>}}{\sigma_{\bz}^2/N}\right)}
\end{align}
where $\bz^{(n)}$ with $n=1,2,\ld,\MCs$ are independent realizations of a zero-mean circularly symmetric complex Gaussian random vector with total variance $\sigma_{\bz}^2$. 
\end{corollary}

\begin{remark}\label{Remark.2}
The results in Theorem~\ref{MD.MC.Theo} are general in the sense it applies to any MD constellation $\mcX$. On the other hand, these results are limited to a circularly symmetric MD AWGN channel. The expressions \eqref{MD.MI.MC} and \eqref{MD.GMI.MC} have been generalized to correlated Gaussian noise in \cite{Sillekens16}. Experimental results for MI and GMI where the circularly symmetric assumption was lifted have been presented in \cite{Eriksson16}.
\end{remark}

\begin{example}[MI for $2$D complex constellations]\label{Example.5}
We computed \eqref{MD.MI.MC} for all the $227$ $2$D complex constellations listed in \cite{Agrell-codes2016}. The obtained results are shown in Fig.~\ref{fig:MIs} (colored lines), where the AWGN capacity is shown as reference. All these curves were computed using $\MCs=10^4$ Monte-Carlo samples and obtained in a few hours on a standard computer. The results in this figure can be used to compare the performance of different modulation formats, even when they have a different number of constellation points. In Fig.~\ref{fig:MIs}, we highlight 64QAM ($M=4096$) and two other constellations that outperform 64QAM: $\mathcal{C}_{4,4096}$ ($M=4096$), and $\mathcal{W}_{4,5698}$ ($M=5698$), the latter proposed in 1974 in \cite{Welti74}. The constellation $\mathcal{W}_{4,5698}$ shows an excellent performance for a very large range of MIs. We warn the reader, however, to be cautious with MI analysis. Constellations that are good in terms of MI might not be good in terms of GMI, as previously shown in \cite{Alvarado2015_JLT} and in Example~\ref{Example.1}. 
\begin{figure}[tbhp]
\centering
  \includegraphics{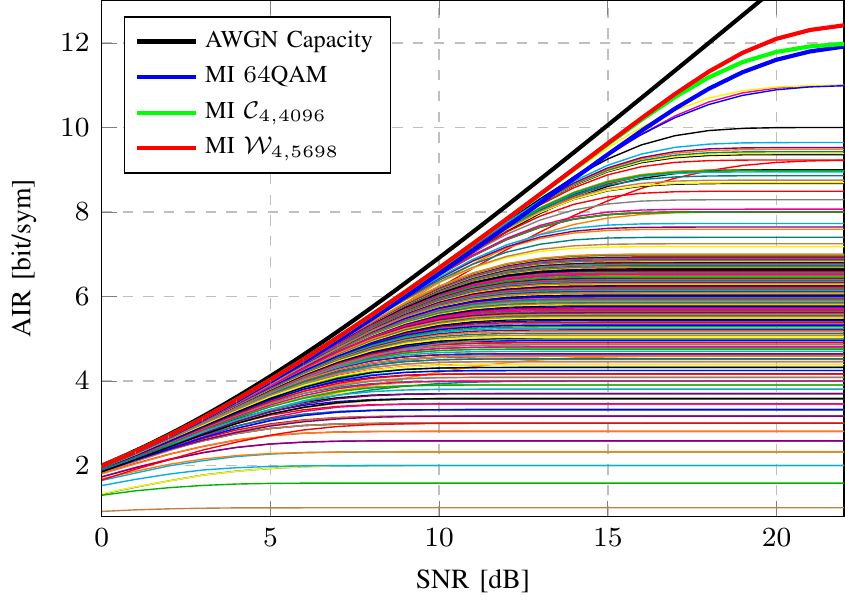}

\caption{MI vs. SNR for all 2D complex (4D real) constellations in \cite{Agrell-codes2016}. The capacity of the AWGN channel in \eqref{C.MD.AWGN} is also shown (thick solid line).}
\label{fig:MIs}
\end{figure}
\end{example}

In the following example, the results in Theorem~\ref{MD.MC.Theo} are particularized to the relevant case of complex constellations.
\begin{example}[Monte-Carlo Complex Constellations]\label{Example.6}
The MI and GMI for the complex AWGN channel can be approximated as
\begin{align}\label{MI.MC}
\nonumber
\MI & \approx m -\frac{1}{M}\sum_{i=1}^{M}\frac{1}{\MCs}\sum_{n=1}^{\MCs} \\
&\qquad \log_{2}  \sum_{j=1}^M \exp{\left(-\frac{|\dij|^{2}+2\Re\set{z^{(n)}\dij}}{\sigma_z^2}\right)}\\
\label{GMI.MC}
\nonumber
\GMI & \approx m-\frac{1}{M}\sum_{k=1}^{m}\sum_{b\in\set{0,1}}\sum_{i\in\mcIkb} \frac{1}{\MCs} \sum_{n=1}^{\MCs} \\
&\qquad\log_{2}\frac{\sum_{p=1}^{M}\exp\left(-\frac{|\dip|^{2}+2\Re\set{z^{(n)} \dip}}{\sigma_z^2}\right)}{\sum_{j\in\mcIkb}\exp\left(-\frac{|\dij|^{2}+2\Re\set{z^{(n)} \dij}}{\sigma_z^2}\right)}
\end{align}
where $z^{(n)}$ with $n=1,2,\ld,\MCs$ are independent realizations of a zero-mean circularly symmetric complex Gaussian random variable with variance $\sigma_z^2$. 
\end{example}

\begin{example}[Estimating GMI from LLRs]\label{Example.7}
The GMI \eqref{GMI.general.LLR} can be estimated via Monte-Carlo integration as \cite[Theorem~4.20]{Alvarado15_Book}
\begin{align}
\label{GMI.general.MC}
\GMI	& \approx m-  \frac{1}{2\MCs}\min_{s\geq 0}\sum_{k=1}^{m}\sum_{b\in\set{0,1}}\sum_{n=1}^{\MCs}\log_{2}\Bigl(1+e^{s(-1)^{b}\lambda^{(n)}_{k,b}}\Bigr)
\end{align}
where $\lambda^{(n)}_{k,b}$, $n=1,2,\ld,\MCs$ are i.i.d. random variables distributed according to the conditional PDF of the L-values $f_{L_{k}|B_{k}}(\lambda|b)$. The minimization over $s$ in \eqref{GMI.general.MC} can be easily approximated (numerically) using the concavity of the GMI in $s$ \cite[Eq.~(4.81)]{Alvarado15_Book}. 
\end{example}

\begin{remark}\label{Remark.3}
The expression in \eqref{GMI.general.MC} is valid for any symbol-wise metric in the form of \eqref{metric.BW}, i.e., for any L-value $L_{k}$ that ignores the dependency between the bits in the symbol. In particular, when the L-values are calculated exactly using \eqref{LLR.general}, the GMI can be estimated using \eqref{GMI.general.MC} and $s=1$, which follows from \cite[Theorem~4.20]{Alvarado15_Book}.
\end{remark}

\begin{example}[Estimating MI in Experiments]\label{Example.8}
In a simulation or experiment where $\MCs M$ symbols were transmitted, the r.h.s. of \eqref{MD.MI.MC} can be estimated in three steps: (i) estimate the noise variance, (ii) for each symbol $\bx_i$, obtain noise realizations $\bz^{(n)}$ by subtracting the transmitted from the received symbols \footnote{After all digital signal processing (filtering, equalization, synchronization, matched filtering, sampling, etc.) such that the conditional sample means of $\by$ are equal to the corresponding $\bx$.} in all the time slots where $\bx_i$ was transmitted, and (iii) use those samples to compute the two innermost sums in \eqref{MD.MI.MC} for all $i=1,\ldots,M$. Although the optical channel is not in general AWGN, this estimated quantity is an AIR using mismatched metrics (more details in Sec.~\ref{Sec:Experiments}).
\end{example}

\begin{example}[Estimating GMI in Experiments]\label{Example.9}
The GMI in \eqref{GMI.MC} (i.e., for the MD AWGN channel) can be estimated following similar steps to those in Example~\ref{Example.8}. For more general memoryless channels, however, the GMI can be estimated using \eqref{GMI.general.MC} as
\begin{align}
\label{GMI.max-log.MC}
\GMI	& \approx m-\frac{1}{\MCs}\min_{s\geq 0}\sum_{k=1}^{m}\sum_{l=1}^{\MCs}\log_{2}\Bigl(1+e^{(-1)^{c_{k,l}}\lambda_{k,l}}\Bigr),
\end{align}
where $\lambda_{k,l}$ are L-values calculated for a given sequences of $m\MCs$ transmitted code bits $c_{k,l}$. This expression holds for any LLR calculation (including max-log LLRs). If the  L-values $\lambda_{k,l}$ are computed via \eqref{LLR.sum-exp.AWGN}, the GMI in \eqref{GMI.max-log.MC} simplifies to
\begin{align}
\label{GMI.sum-exp.MC}
\GMI	& \approx m-\frac{1}{\MCs}\sum_{k=1}^{m}\sum_{l=1}^{\MCs}\log_{2}\Bigl(1+e^{(-1)^{c_{k,l}}\lambda_{k,l}}\Bigr).
\end{align}
\end{example}

\begin{remark}\label{Remark.4}
Note again that to calculate the GMI for max-log LLRs (or more generally, for mismatched LLRs), \eqref{GMI.max-log.MC} should be used. Using \eqref{GMI.sum-exp.MC} for mismatched LLRs will result in a rate lower than the true one. In other words, the minimization over $s$ in \eqref{GMI.max-log.MC} is a mandatory step for an information-theoretically precise treatment of GMI with mismatched LLRs.
\end{remark}

\subsection{Gauss-Hermite Quadrature}\label{Sec:GH}

For any real-valued function $g(r):\mathbb{C}\to\mathbb{R}$ with bounded $2\GHs$-th derivative, the $\GHs$-point Gauss-Hermite quadrature is given by \cite[Sec.~7.3.4]{Churchhouse81_Book}
\begin{align}\label{GH_C}
\int_{\mathbb{C}}\exp{(-|r|^2)}g(r)\,\tnr{d}r \approx \sum_{l_1=1}^{\GHs}\alpha_{l_1}\sum_{l_2=1}^{\GHs}\alpha_{l_2} g(\xi_{l_1}+\jmath\xi_{l_2})
\end{align}
where the quadrature nodes $\xi_l$ 
and the weights $\alpha_l$
can be easily found (numerically) for different values of $\GHs$. This value determines the trade-off between the computation speed and the accuracy of the quadrature. The approximation \eqref{GH_C} is exact when $\GHs\rightarrow\infty$. Table~\ref{tab:GH} shows the quadrature nodes and weights for $J=10$.

\begin{table}[t]
\caption{Quadrature nodes $\xi_l$ and weights $\alpha_l$ for $J=10$}
\centering
 \renewcommand{\arraystretch}{1.1}
\begin{tabular}{ccc}
\hline
$l$ & $\xi_l$ & $\alpha_l$ \\
\hline
1 &	-3.4362	& 0.000007640
   \\
2 &-2.5327	& 0.001343645
   \\
3 &-1.7567	& 0.033874394
   \\
4 &-1.0366	& 0.240138611
   \\
5 &-0.3429	& 0.610862633
   \\
6 &0.3429		& 0.610862633
   \\
7 &1.0366		& 0.240138611
   \\
8 &1.7567		& 0.033874394
   \\
9 &2.5327		& 0.001343645
   \\
10 &	3.4362 	& 0.000007640\\
\hline
\end{tabular}
\label{tab:GH}
\end{table}

A straightforward (but not unique) generalization of \eqref{GH_C} to MD integrals is 
\begin{align}\label{GH_C.MD}
\int_{\mathbb{C}^N}&\exp{(-\|\br\|^2)}g(\br)\,\tnr{d}\br \approx  \sum_{l_1=1}^{\GHs}\sum_{l_2=1}^{\GHs}\ldots\sum_{l_{2N}=1}^{\GHs} g(\boldsymbol{\xi})\prod_{n=1}^{2N} \alpha_{l_n}
\end{align}
where in this case $g(\br):\mathbb{C}^N\to\mathbb{R}$, and the vector of nodes
\begin{align}\label{xi.vec}
\bxi\triangleq [\xi_{l_1}+\jmath\xi_{l_2},\xi_{l_3}+\jmath\xi_{l_4},\ldots,\xi_{l_{2N-1}}+\jmath\xi_{l_{2N}}].
\end{align}

The following corollary gives Gauss-Hermite approximations of the MI and GMI for the MD AWGN channel. 
\begin{corollary}[Gauss-Hermite Approximations]\label{MD.GH.Theo}
The MI and GMI for an MD AWGN channel can be approximated as
\begin{align}
\label{MI.general.8}
\MI       & \approx m-\frac{1}{M\pi^N}\sum_{i=1}^{M}
 \sum_{l_1=1}^{\GHs}\sum_{l_2=1}^{\GHs}\ldots\sum_{l_{2N}=1}^{\GHs} g_i^\MI(\boldsymbol{\xi})\prod_{n=1}^{2N} \alpha_{l_n}
\end{align}
with
\begin{align}\label{MI.MC.MD.gi}
g_i^\MI(\boldsymbol{\xi}) = \log_{2}{\sum_{j=1}^{M}\exp{\left(-\frac{\|\bdij\|^2+2\sigma_{\bz}/\sqrt{N}\Re\set{\left<\bxi,\bdij\right>}}{\sigma_{\bz}^{2}/N}\right)}}
\end{align}
and
\begin{align}
\nonumber
\GMI       & \approx m- \frac{1}{M\pi^N}\sum_{k=1}^{m}\sum_{b\in\set{0,1}}\sum_{i\in\mcIkb} \\
\label{GMI.general.8}
&\qquad\qquad\qquad\qquad
 \sum_{l_1=1}^{\GHs}\sum_{l_2=1}^{\GHs}\ldots\sum_{l_{2N}=1}^{\GHs} g_i^\GMI(\boldsymbol{\xi})\prod_{n=1}^{2N} \alpha_{l_n}
\end{align}
with
\begin{align}\label{GMI.MC.MD.gi}
g_i^\GMI(\boldsymbol{\xi}) =
\log_{2}\frac{\sum_{p=1}^{M}\exp\left(-\frac{\|\bdip\|^{2}+2\sigma_{\bz}/\sqrt{N}\Re\set{\left<\bxi,\bdip\right>}}{\sigma_{\bz}^{2}/N}\right)}{\sum_{j\in\mcIkb}\exp\left(-\frac{\|\bdij\|^{2}+2\sigma_{\bz}/\sqrt{N}\Re\set{\left<\bxi,\bdij\right>}}{\sigma_{\bz}^{2}/N}\right)}
\end{align}
where $\bxi$ is given by \eqref{xi.vec}.
\end{corollary}
\begin{IEEEproof}
See Appendix~\ref{Appendix.MD.GH.Theo}.
\end{IEEEproof}

\begin{example}
The MI and GMI for the complex AWGN channel is
\begin{align}
\label{MI.GH.Complex.1}
\nonumber
 \MI       & \approx m-\frac{1}{M\pi}\sum_{i=1}^{M}\sum_{l_1=1}^{\GHs}\alpha_{l_1}\sum_{l_2=1}^{\GHs}\alpha_{l_2}  \cd\\
 & \log_{2}{\sum_{j=1}^{M}\exp{\left(-\frac{|\dij|^2+2\sigma_{z}\Re\set{(\xi_{l_1}+\jmath\xi_{l_2}) \dij}}{\sigma_{z}^{2}}\right)}},\\
\label{MI.GH.Complex.2}
\nonumber
\GMI       & \approx m-  \frac{1}{M\pi}\sum_{k=1}^{m}\sum_{b\in\set{0,1}}\sum_{i\in\mcIkb}\sum_{l_1=1}^{\GHs}\alpha_{l_1}\sum_{l_2=1}^{\GHs}\alpha_{l_2} \cd \\
&\log_{2}\frac{\sum_{p=1}^{M}\exp\left(-\frac{|\dip|^{2}+2\sigma_{z}\Re\set{(\xi_{l_1}+\jmath\xi_{l_2})\dip}}{\sigma_{z}^{2}}\right)}{\sum_{j\in\mcIkb}\exp\left(-\frac{|\dij|^{2}+2\sigma_{z}\Re\set{(\xi_{l_1}+\jmath\xi_{l_2})\dij}}{\sigma_{z}^{2}}\right)}.
\end{align}
These expressions are obtained by particularizing Corollary~\ref{MD.GH.Theo} to the case of $N=1$ complex dimensions and were used to obtain the curves in Example~\ref{Example.1} in Sec.~\ref{Sec:General:Applications} with $J=20$.
\end{example}

\begin{remark}\label{Remark.5}
The advantage of Gauss-Hermite quadrature over Monte-Carlo is that the former is normally much faster for real and two-dimensional (complex) constellations (see \cite[Example 4.26]{Alvarado15_Book}). Monte-Carlo integration is better suited for constellations with more dimensions or when the channel law is unknown, as discussed in \secref{Sec:Experiments}. Another advantage of Gauss-Hermite quadrature is that its evaluation does not depend on random samples (as in the case of Monte-Carlo), and thus, it is better-suited for numerical optimizations.\footnote{The randomness of the Monte-Carlo estimates in a numerical optimization routine can, however, be avoided using the same random seed.}
\end{remark}

\begin{example}[MI for $\infty$-QAM]\label{Example.11}
Fig.~\ref{fig:MIsDense} shows the MI for $M$QAM with cardinalities up to $M\leq 65536=2^{16}$, i.e., for constellations with maximum spectral efficiency of  $32$~bit/sym (sometimes called $\infty$-QAM constellations). These results were obtained via \eqref{MI.GH.Complex.1} with $J=10$ and highlight how Gauss-Hermite quadrature is a very powerful tool for constellations with a small number of dimensions. At high SNR, the gap between the MI envelope and the capacity of the MD AWGN channel reaches the ``ultimate shaping gain'' 1.53~dB \cite{Forney1992TransIT_TrellisShaping}.

\begin{figure}[tbhp]
\centering 
  \includegraphics{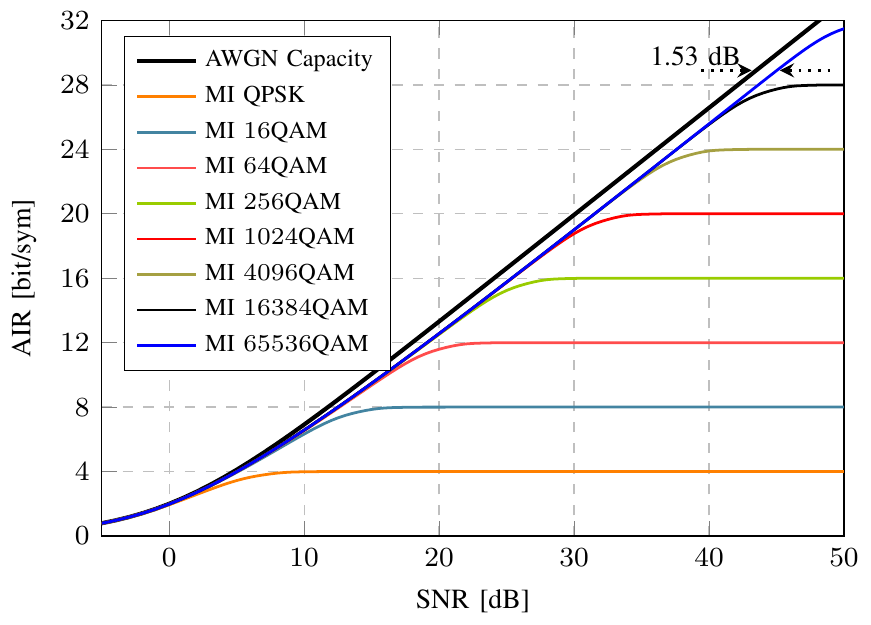}
\caption{MI vs. SNR for dense 2D complex $M$QAM constellations: $M\leq 65536=2^{16}$, i.e., maximum spectral efficiency of  $32$~bit/sym. The capacity of the AWGN channel in \eqref{C.MD.AWGN} is also shown (thick solid line).}
\label{fig:MIsDense}
\end{figure}

\end{example}

\section{MI and GMI for the Nonlinear Optical Channel}\label{Sec:Experiments}

The fiber-optical channel is a nonlinear channel with memory, and thus, the AIRs we discussed in the previous sections (MI and GMI) are (possibly loose) lower bounds on the rates that can be achieved when memory and fiber nonlinearities are taken into account. After a very brief discussion of the capacity of such a channel, the majority of this section is devoted to lower bounds on capacity (i.e., AIRs) for the memoryless demapper shown in Fig.~\ref{channel_blockdiagram}. Ready-to-use expression to compute these AIRs lower bounds are presented.

\subsection{Channel Capacity}

The nonlinear optical channel (where the propagation of the optical field is governed by the nonlinear Schr\"{o}dinger equation) is a channel with memory. The channel law can therefore be described by the conditional PDF $f_{\un{\bY}|\un{\bX}}(\un{\bY}|\un{\bX})$, as shown in Fig.~\ref{channel_blockdiagram}.

For channels with memory, and under certain assumptions on information stability \cite[Sec.~I]{Verdu94}, the channel capacity is
\begin{align}
\label{C.memory}
C^{\tnr{mem}} = \lim_{l\rightarrow\infty} \sup_{p_{\un{\bX}}} \frac{1}{l} I(\un{\bX};\un{\bY}),
\end{align}
where $\un{\bX} = [\bX_1,\bX_{2},\ldots,\bX_l]$, $\un{\bY} = [\bY_1,\bY_{2},\ldots,\bY_l]$, $I(\un{\bX};\un{\bY})$ is defined as a multidimensional MI analogous to \eqref{MI.def}, and the maximization is over all distributions of the random \emph{vectors} $\un{\bX}$ satisfying a power constraint.

\begin{remark}
The capacity expression in \eqref{C.memory} is based on a discrete-time channel model described by $f_{\un{\bY}|\un{\bX}}(\un{\bY}|\un{\bX})$, and thus, it has units of bit per symbol (or bit per channel use). This capacity expression does consider memory, yet it does not take into account bandwidth (or potential bandwidth expansion at high powers). Therefore, this analysis cannot be straightforwardly used to determine the capacity in bit/s/Hz. More details on this issue can be found in \cite[Sec.~V]{Kramer2016} \cite{Kramer2017}.
\end{remark}

For independent and identically distributed symbols $\bX$ and a suboptimal receiver that neglects any residual memory after DSP (i.e., for the demapper in Fig.~\ref{channel_blockdiagram}), the memoryless MI \eqref{MI.def} becomes the largest achievable rate. Such a setup is shown in \figref{channel_blockdiagram} by the information-theoretic channel. Furthermore, the capacity in \eqref{C.memory} is lower-bounded by 
\begin{align}\label{C.bounds}
C^{\tnr{mem}}\geq \MI \geq \GMI.
\end{align}

\subsection{Lower Bounds on the MI and GMI}

The MI for the (suboptimal) receiver described above and independent and uniformly distributed symbols is given by \eqref{MI.general.0}. As shown in \eqref{C.bounds}, this also gives a lower bound on the capacity of the channel with memory in \eqref{C.memory}. We are therefore interested in computing \eqref{MI.general.2} with $g_{i}^\MI(\by)$ given by \eqref{MI.gi}. For the nonlinear optical channel, however, the channel law $f_{\bY|\bX}$ is usually not known analytically. The MI can be lower bounded as \cite[Sec.~VI]{Arnold2006TransIT_AchievableRates}
\begin{align}\label{MI.LB}
I \geq \tilde{I} = m+\frac{1}{M}\sum_{i=1}^{M} \int_{\mathds{C}^N}f_{\bY|\bX}(\by|\bx_i) \tilde{g}_i^\MI(\by)\, \tnr{d}\by,
\end{align}
where
\begin{align}\label{MI.gi.q}
\tilde{g}_i^\MI(\by) = \log_{2}\frac{q_{\bY|\bX}(\by|\bx_i)}{\sum_{j=1}^{M}q_{\bY|\bX}(\by|\bx_j)},
\end{align}
and $q_{\bY|\bX}$ is any PDF (usually called the auxiliary channel). In particular, if $q_{\bY|\bX}$ is the true channel law $f_{\bY|\bX}$, \eqref{MI.LB} holds with equality. Clearly, the more ``close'' $q_{\bY|\bX}$ is to $f_{\bY|\bX}$, the tighter the bound in \eqref{MI.LB}.

A common choice for $q_{\bY|\bX}$ is a circularly symmetric Gaussian PDF with variance $\sigma_{\bz}^{2}$. The circularly symmetric Gaussian PDF assumption might be suboptimal, but in the absence of a better (non-Gaussian) model, the circularly symmetric Gaussian noise assumption is reasonable.\footnote{It is experimentally demonstrated in \cite{Eriksson16} that for long-haul dispersion-unmanaged optical systems a Gaussian assumption for the noise is a very good approximation.} For this auxiliary channel, and in analogy to \eqref{MI.general.3}, the lower bound \eqref{MI.LB} becomes
\begin{align}
\label{MI.general.3.q}
\nonumber
\tilde{I}  & = m-\frac{1}{M}\sum_{i=1}^{M} \int_{\mathds{C}^N}f_{\bY|\bX}(\by|\bx_i)\cd\\
& \quad \log_{2}{\sum_{j=1}^{M}\exp{\left(-\frac{\|\bdij\|^2+2\Re\set{\left<\by-\bx_{i},\bdij\right>}}{\sigma_{\bz}^2/N}\right)}} \, \tnr{d}\by
\end{align}
where we again dropped the complex conjugate in the term $\by-\bx_{i}$ due to the circular symmetry of the PDF $q_{\bY|\bX}$.

In analogy to \eqref{MI.general.3.q}, a lower bound on the GMI in \eqref{MD.GMI.general.5} can be expressed as
\begin{align}
\label{GMI.MC.q}
\nonumber
\GMI & \geq \tilde{\GMI} =m-\frac{1}{M}\sum_{k=1}^{m}\sum_{b\in\set{0,1}}\sum_{i\in\mcIkb} \int_{\mathds{C}^N}f_{\bY|\bX}(\by|\bx_i) \cd \\
& \qquad \,\, \log_{2}\frac{\sum_{p=1}^{M}\exp\left(-\frac{\|\bdip\|^{2}+2\Re\set{\left<\by-\bx_{i},\bdip\right>}}{\sigma_{\bz}^2/N}\right)}{\sum_{j\in\mcIkb}\exp\left(-\frac{\|\bdij\|^{2}+2\Re\set{\left<\by-\bx_{i},\bdij\right>}}{\sigma_{\bz}^2/N}\right)}\, \tnr{d}\by
\end{align}

In view of \eqref{MC}, the MI and GMI lower bounds in \eqref{MI.general.3.q} and \eqref{GMI.MC.q} can be approximated via Monte-Carlo integration as
\begin{align}
\label{MI.general.3.q.MC}
\nonumber
\tilde{I}  & \approx m-\frac{1}{M}\sum_{i=1}^{M} \frac{1}{\MCs}\sum_{n=1}^{\MCs} \\
&\quad \log_{2}{\sum_{j=1}^{M}\exp{\left(-\frac{\|\bdij\|^2+2\Re\set{\bigl<\bs_{i}^{(n)}-\bx_{i},\bdij\bigr>}}{\sigma_{\bz}^2/N}\right)}}
\end{align}
and
\begin{align}
\nonumber
\label{GMI.MC.q.MC}
\tilde{\GMI} & \approx m-\frac{1}{M}\sum_{k=1}^{m}\sum_{b\in\set{0,1}}\sum_{i\in\mcIkb} \frac{1}{\MCs} \sum_{n=1}^{\MCs} \\
&\qquad \,\log_{2}\frac{\sum_{p=1}^{M}\exp\left(-\frac{\|\bdip\|^{2}+2\Re\set{\bigl<\bs_{i}^{(n)}-\bx_{i},\bdip\bigr>}}{\sigma_{\bz}^2/N}\right)}{\sum_{j\in\mcIkb}\exp\left(-\frac{\|\bdij\|^{2}+2\Re\set{\bigl<\bs_{i}^{(n)}-\bx_{i},\bdij\bigr>}}{\sigma_{\bz}^2/N}\right)}
\end{align}
where $\bs_{i}^{(n)}$ with $n=1,2,\ld,\MCs$ are samples taken from the channel conditioned on that the transmitted symbol was $\bX=\bx_{i}$, i.e., samples from the (analytically unknown) PDF $f_{\bY|\bX}(\by|\bx_i)$. Note that the expressions in \eqref{MI.general.3.q.MC} and \eqref{GMI.MC.q} are not lower bounds, but only an approximations to the lower bounds. As the number of samples $\MCs$ tends to infinity, however, these approximations tend to the true bounds in \eqref{MI.general.3.q} and \eqref{GMI.MC.q}.

%

\section{Conclusions}

In this paper, we reviewed, achievable information rates for fiber optical communication systems and showed that they are versatile and powerful design metrics for coded systems. Different ready-to-use approximations were presented and discussed. The main focus of this paper was on the multidimensional AWGN channel. This model is very relevant to current optical transceivers used in long-haul dispersion-uncompensated links where a memoryless demapper is used.

The methods described in this paper can be generalized to non-AWGN channels and also to channels with memory. This methodology could for example be used to design multidimensional constellations and coded modulation tailored to the nonlinear optical channel. This is left for further investigation. 
All the analysis presented here was also based on infinite-block length assumptions and universal codes. Both finite-block length analyses and analysis of code universality are interesting future research avenues.

\section*{Acknowledgments}\label{Sec:Ack}

The authors would like to thank Dr. Gabriele Liga (University College London, UK) for providing some of the data in Fig.~\ref{Reach} (previously published in \cite{Liga16}) as well as to Dr. Laurent Schmalen (Nokia Bell Labs, Germany) for providing the data in Fig.~\ref{predictionNB}, previously reported in \cite{Schmalen17}.

\appendices

\section{Proof of Theorem~\ref{MI.MDAWGN.Theo}}\label{Appendix.MI.MDAWGN.Theo}
Using \eqref{MD.AWGN.PDF} in \eqref{MI.gi}, we obtain
\begin{align}
\label{MI.gi.AWGN.0}
g_i^\MI(\by)  & = \log_{2}\frac{\exp{\left(-\frac{\|\by-\bx_i\|^2}{\sigma_{\bz}^2/N}\right)}}{\sum_{j=1}^{M}\exp{\left(-\frac{\|\by-\bx_j\|^2}{\sigma_{\bz}^2/N}\right)}}\\
\label{MI.gi.AWGN.1}
        & = \log_{2}\frac{\exp{\left(-\frac{\|\bz_i(\by)\|^2}{\sigma_{\bz}^2/N}\right)}}{\sum_{j=1}^{M}\exp{\left(-\frac{\|\bz_i(\by)+\bdij\|^2}{\sigma_{\bz}^2/N}\right)}}\\
\label{MI.gi.AWGN.2}
        &  = -\log_{2}{\sum_{j=1}^{M}\exp{\left(-\frac{\|\bdij\|^2+2\Re\set{\left<\bz_i^*(\by),\bdij\right>}}{\sigma_{\bz}^2/N}\right)}},
\end{align}
where $\bz_i(\by)=\by-\bx_i$, $\left<\bz_1,\bz_2\right>$ is the inner product, and \eqref{MI.gi.AWGN.2} follows by using $\|\bz_1+\bz_2\|^2=\|\bz_1\|^2+\|\bz_2\|^2+2\Re\set{\left<\bz_1^*,\bz_2\right>}$, where $\bz^*$ denotes complex conjugate\footnote{This can be proven using the identity $\|\bz\|^2=\bz^{H}\bz$, where $(\cd)^{H}$ is Hermitian transpose.}.

The expression in \eqref{MI.gi.AWGN.2} is used in \eqref{MI.general.2} to obtain
\begin{align}
\label{MI.general.3}
\nonumber
I  & = m-\frac{1}{M}\sum_{i=1}^{M} \int_{\mathds{C}^N}f_{\bY|\bX}(\by|\bx_i)\cd \\ 
	&\,\,\quad \log_{2}{\sum_{j=1}^{M}\exp{\left(-\frac{\|\bdij\|^2+2\Re\set{\left<\bz_i(\by),\bdij\right>}}{\sigma_{\bz}^2/N}\right)}}\, \tnr{d}\by\\
\label{MI.general.4}
\nonumber
        & = m-\frac{1}{M}\sum_{i=1}^{M} \int_{\mathds{C}^N}f_{\bZ_i}(\bz_i) \cd \\
        & \,\,\quad \log_{2}{\sum_{j=1}^{M}\exp{\left(-\frac{\|\bdij\|^2+2\Re\set{\left<\bz_i,\bdij\right>}}{\sigma_{\bz}^2/N}\right)}}\, \tnr{d}\bz_i\\
\label{MI.general.5}
\nonumber
        & = m-\frac{1}{M}\sum_{i=1}^{M} \int_{\mathds{C}^N}f_{\bZ}(\bz) \cd \\
        & \,\,\quad \log_{2}{\sum_{j=1}^{M}\exp{\left(-\frac{\|\bdij\|^2+2\Re\set{\left<\bz,\bdij\right>}}{\sigma_{\bz}^2/N}\right)}}\, \tnr{d}\bz
\end{align}
where \eqref{MI.general.4} follows from integrating over $\bz_i$ instead of over $\by$. Given that $\bZ_i$ is a circularly symmetric zero-mean complex Gaussian random variable with variance $\sigma_z^2$, the complex conjugate of $\bz_i^*(\by)$ in \eqref{MI.general.3} was  dropped. Furthermore, because the statistics of $\bZ_{i}$ do not depend on $i$, this subindex can be dropped, which gives \eqref{MD.MI.general.5}, and thus, completes the proof.

\section{Proof of Theorem~\ref{GMI.MDAWGN.Theo}}\label{Appendix.GMI.MDAWGN.Theo}

The proof follows similar steps to that in Appendix~\ref{Appendix.MI.MDAWGN.Theo}. In particular, \eqref{MD.AWGN.PDF} is used to express $g_i^\GMI(\by)$ in \eqref{GMI.gi} as
\begin{align}
\label{GMI.gi.AWGN.2}
g_i^\GMI(\by)  & = -\log_{2}{\frac{\sum_{p=1}^{M}\exp{\left(-\frac{\|\bdip\|^2+2\Re\set{\left<\bz_i^*(\by),\bdip\right>}}{\sigma_{\bz}^2/N}\right)}}{\sum_{j\in\mcIkb}\exp{\left(-\frac{\|\bdij\|^2+2\Re\set{\left<\bz_i^*(\by),\bdij\right>}}{\sigma_{\bz}^2/N}\right)}}}.
\end{align}
Both the index $i$ and the complex conjugate in $\bz_i^*(\by)$ are dropped and the resulting expression is used in \eqref{GMI.Integral.3} to obtain \eqref{MD.GMI.general.5}.


\section{Proof of Corollary~\ref{MD.GH.Theo}}\label{Appendix.MD.GH.Theo}

The MI and GMI in \eqref{MD.MI.general.5} and \eqref{MD.GMI.general.5} can be expressed as
\begin{align}
\label{MI.general.6}
\nonumber
 I       & = m-\frac{1}{M}\sum_{i=1}^{M}\int_{\mathds{C}^N} \frac{1}{(\pi\sigma_{\bz}^2/N)^N}\exp{\left(-\frac{\|\bz\|^2}{\sigma_{\bz}^2/N}\right)} \cd\\
 & \qquad \log_{2}{\sum_{j=1}^{M}\exp{\left(-\frac{\|\bdij\|^2+2\Re\set{\left<\bz,\bdij\right>}}{\sigma_{\bz}^2/N}\right)}} \, \tnr{d}\bz,
\end{align}
and
\begin{align}
\label{GMI.general.6}
\nonumber
\GMI       & = m-  \frac{1}{M}\sum_{k=1}^{m}\sum_{b\in\set{0,1}}\sum_{i\in\mcIkb}\int_{\mathds{C}^N}\frac{1}{(\pi\sigma_{\bz}^2/N)^N}\exp{\left(-\frac{\|\bz\|^2}{\sigma_{\bz}^2/N}\right)}  \cd \\
& \qquad \log_{2}\frac{\sum_{p=1}^{M}\exp\left(-\frac{\|\bdip\|^{2}+2\Re\set{\left<\bz,\bdip\right>}}{\sigma_{\bz}^2/N}\right)}{\sum_{j\in\mcIkb}\exp\left(-\frac{\|\bdij\|^{2}+2\Re\set{\left<\bz,\bdij\right>}}{\sigma_{\bz}^2/N}\right)}\, \tnr{d}\bz,
\end{align}
respectively. Using the change of variables $\bz=\sqrt{\frac{\sigma_{\bz}^2}{N}} \br$, we obtain
\begin{align}
\label{MI.general.7}
\nonumber
 I       & = m-\frac{1}{M\pi^N}\sum_{i=1}^{M}\int_{\mathds{C}^N}\exp{\left(-\|\br\|^2\right)}  \cd\\
 &  \log_{2}{\sum_{j=1}^{M}\exp{\left(-\frac{\|\bdij\|^2+2\sigma_{\bz}/\sqrt{N}\Re\set{\left<\br,\bdij\right>}}{\sigma_z^{2}/N}\right)}} \, \tnr{d}\br,
\end{align}
and
\begin{align}
\label{GMI.general.7}
\nonumber
\GMI       & = m-  \frac{1}{M\pi^N}\sum_{k=1}^{m}\sum_{b\in\set{0,1}}\sum_{i\in\mcIkb} \int_{\mathds{C}^N}\exp{\left(-\|\br\|^2\right)}\cdot \\
	& \log_{2}\frac{\sum_{p=1}^{M}\exp\left(-\frac{\|\bdip\|^{2}+2\sigma_{\bz}/\sqrt{N}\Re\set{\left<\br,\bdip\right>}}{\sigma_{\bz}^{2}/N}\right)}{\sum_{j\in\mcIkb}\exp\left(-\frac{\|\bdij\|^{2}+2\sigma_{\bz}/\sqrt{N}\Re\set{\left<\br,\bdij\right>}}{\sigma_{\bz^{2}}/N}\right)}\, \tnr{d}\br.
\end{align}
The proof is completed by applying \eqref{GH_C.MD} to \eqref{MI.general.7} and \eqref{GMI.general.7}.

\balance

\bibliographystyle{IEEEtran}
\bibliography{myJournalsFull,references,myPublications}

\end{document}